\documentclass[final,authoryear,5p,times,twocolumn]{elsarticle}

\usepackage{graphicx,epstopdf,amsfonts,amsmath,amssymb,array,mathrsfs}

\usepackage{color}

\usepackage{fancyhdr}

\usepackage{lineno}

\usepackage{tabu,multirow}

\newcommand{\vince}[1]{{\color{black}{#1}}}

\newcommand{\abbreviations}[1]{%
  \nonumnote{\textit{Abbreviations:\enspace}#1}}

\journal{NeuroImage}

  \makeatletter
  \def\ps@pprintTitle{%
  \let\@oddhead\@empty \let\@evenhead\@empty
 \def\@oddfoot{\leftline{\footnotesize \sl Article published in \textbf{NeuroImage 270 (2023) 119953}. DOI: 10.1016/j.neuroimage.2023.119953. Open access under CC BY-NC-ND license.}}
  \let\@evenfoot\@empty
  }
  \makeatother

\begin{document}


\begin{frontmatter}



\title{{\bf Exploring the limits of MEG spatial resolution with multipolar expansions}}

\author[lcfc,meg]{Vincent Wens\corref{vw}}

\address[lcfc]{LN$^2$T -- Laboratoire de Neuroanatomie et Neuroimagerie translationnelles, UNI -- ULB Neuroscience Institute, Universit\'e libre de Bruxelles (ULB), Brussels, Belgium}
\address[meg]{Department of Translational Neuroimaging, H.U.B. -- H\^opital Erasme, Brussels, Belgium}

\cortext[vw]{Corresponding author. \textit{Address:} Department of Translational Neuroimaging, H.U.B. -- H\^opital Erasme, 808 route de Lennik, 1070 Brussels, Belgium. \textit{E-mail address:} \texttt{vincent.wens@ulb.be}.}

\abbreviations{ECoG, electrocorticography; EEG, electroencephalography; MEG, magnetoencephalography; MRI, magnetic resonance imaging; OPM, \vince{optically pumped} magnetometer; QZFM, quantum zero-field magnetometer; SQUID, superconducting quantum interference device; SNR, signal-to-noise ratio.}


\begin{abstract}

\noindent The advent of scalp magnetoencephalography (MEG) based on optically pumped magnetometers (OPMs) may represent a step change in the field of human electrophysiology. Compared to cryogenic MEG based on superconducting quantum interference devices (SQUIDs, placed $2$--$4$ cm above scalp), scalp MEG promises significantly higher spatial resolution imaging but it also comes with numerous challenges regarding how to optimally design OPM arrays. In this context, we sought to provide a systematic description of MEG spatial resolution as a function of the number of sensors (allowing comparison of low- vs.~high-density MEG), sensor-to-brain distance (cryogenic SQUIDs vs.~scalp OPM), sensor type (magnetometers vs.~gradiometers; single- vs.~multi-component sensors), and signal-to-noise ratio. To that aim, we present an analytical theory based on MEG multipolar expansions that enables, once supplemented with experimental input and simulations, quantitative assessment of the limits of MEG spatial resolution in terms of two qualitatively distinct regimes. In the regime of asymptotically high-density MEG, we provide a mathematically rigorous description of how magnetic field smoothness constraints spatial resolution to a slow, logarithmic divergence. In the opposite regime of low-density MEG, it is sensor density that constraints spatial resolution to a  faster increase following a square-root law. The transition between these two regimes controls how MEG spatial resolution saturates as sensors approach sources of neural activity. This two-regime model of MEG spatial resolution integrates known observations (e.g., the difficulty of improving spatial resolution by increasing sensor density, the gain brought by moving sensors on scalp, or the usefulness of multi-component sensors) and gathers them under a unifying theoretical framework that highlights the underlying physics and reveals properties inaccessible to simulations. We propose that this framework may find useful applications to benchmark the design of future OPM-based scalp MEG systems.

\end{abstract}


\begin{keyword}

High-density MEG; Magnetic field smoothness; Magnetoencephalography; Optically pumped magnetometry; Scalp MEG.

\medskip
\emph{Highlights:}

\ \ \ \ \ \ \ \ \  $\bullet$ We develop a two-regime theory describing the limits of MEG spatial resolution. \\
\ \ \ \ \ \ \ \ \  $\bullet$ The low-density regime exhibits the advantage of multi-component MEG sensors. \\
\ \ \ \ \ \ \ \ \  $\bullet$ The high-density regime reveals a slow divergence as sensors are added to MEG. \\
\ \ \ \ \ \ \ \ \  $\bullet$ Scalp MEG exhibits saturated resolution through an interplay of the two regimes. \\
\ \ \ \ \ \ \ \ \  $\bullet$ This theoretical framework may be helpful to design new generation scalp MEG. \\

\end{keyword}

\end{frontmatter}




\pagestyle{fancy}
\fancyhead{}
\fancyhead[L]{\sl {Exploring the limits of MEG spatial resolution with multipolar expansions}}
\fancyhead[R]{\sl {V.~Wens (2023)}}
\renewcommand{\headrulewidth}{0pt}


\section{Introduction}\label{Introduction}

The physics of electric and magnetic fields sets fundamental limits to the spatial resolution of non-invasive electrophysiology.
As these fields spread from the brain to extra-cranial sensors \citep[scalp electrodes for electroencephalography, EEG; magnetometers and gradiometers for magnetoencephalography, MEG; see, e.g.,][]{Hamalainen1993MEG}, fine details of neural current source distributions get blurred and information is lost. 
This loss of spatial resolution and the accompanying decrease in field amplitude lie at the heart of the MEG/EEG inverse problem \citep{Hamalainen1993MEG} and thus cannot be overcome fully by technological developments \citep{Tarantola2006Nature}. Still, our ability to harvest smaller and smaller details of brain electrophysiological signals improves as technology evolves.

The development of scalp MEG based on optically pumped magnetometers (OPMs) may lead to major advances in this regard \citep{Boto2018}. By avoiding the heavy cryogenics needed for MEG systems based on superconducting quantum interference devices (SQUIDs), this technology allows to place magnetometers closer to the scalp (from about 2--4 cm above scalp for SQUIDs to about 5 mm for OPMs) and thus leads to substantial improvements in signal focality and amplitude \citep{Boto2016, Feys2022, Ilvanainen2017, Iivanainen2020}. The ensuing refinement in MEG data remains limited in practice due to the relatively low number of magnetometers in current wearable OPM systems \citep[up to 50 in][]{Hill2020, Boto2021} and their sensitivity to environmental noise \citep{Boto2018, Seymour2022}. The situation is evolving rapidly though, as denser OPM arrays \citep{Boto2016, Ilvanainen2017}, new types of OPM sensors \citep{Labyt2019, Borna2020, Nardelli2020, Brookes2021}, and background field cancellation techniques \citep{Holmes2018, Holmes2019, IIVANAINEN2019, Mellor2022} are being invented.
In this exciting context, one important question to consider is how much detail of the neuromagnetic field can be harvested by MEG sensor arrays \emph{if} we could place as many sensors as we wanted on these arrays. In other words, what are the limits of MEG spatial resolution, given a system design (i.e., sensor coverage, density, sensor-to-brain distance, field sensitivity, and noise level)? This question is admittedly abstract, since practical constraints such as sensor size (about $2\, \textrm{cm}^2$ scalp contact area for state-of-the-art OPMs) or cost restrict the number of available sensors in current MEG arrays. That said, its answer could have an important impact on the development of future OPM systems as sensors become smaller and cheaper. Characterising the limits of MEG spatial resolution quantitatively and systematically would allow to assess how many sensors are ideally needed to map neuromagnetic fields as precisely as possible, and how this number is affected by system design.

This type of question was already asked in the early days of the whole-brain-covering SQUID array technology. In their seminal study, \cite{Ahonen1993} used a two-dimensional version of Nyquist's sampling theorem to estimate the distribution of radial magnetic sensors needed to image dipolar magnetic fields faithfully without aliasing. This information was crucial for the development of modern multi-SQUID systems. More recent approaches in the context of OPM developments focused instead on simulated models of MEG signals. Forward modeling (i.e., the explicit numerical evaluation of field propagation from neural current sources to sensors) was used extensively to estimate the resolution gain expected when passing from SQUIDs to OPMs \citep[i.e., with similar design but placed on scalp; see][]{Boto2016, Ilvanainen2017, Tierney2020}. 
Multipolar expansions \citep{Jackson,Zangwill2012} provide another modeling technique that is ideally suited to investigate spatial resolution as they decompose MEG data in terms of angular frequency, i.e., a measure of spatial scale on sensor topographies. 
This decomposition underlies signal-space separation, a preprocessing technique of SQUID signals that allows to suppress both focal sensor noise at high angular frequency and widespread long-distance environmental magnetic interferences at low angular frequency \citep{Taulu2004, Taulu2005}.  \cite{Tierney2022} explored OPM spatial sampling with simulations built from MEG multipolar expansions. However, simulation-based approaches are impractical to handle the case of asymptotically dense sensor arrays needed to assess the limits of MEG spatial resolution.

Here, we expand upon the multipolar expansion technique and provide a systematic framework for MEG spatial resolution that encompasses its limits. We use an analytical description of neuromagnetic field smoothness in asymptotically high-density MEG with hemispherical geometry to characterize spatial resolution in terms of the highest angular frequency accessible to the array.
Given that magnetic field spread exerts a smoothing on extra-cranial neuromagnetic topographies, we hypothesized that MEG spatial resolution would converge to a definite limit controlled by field smoothness once sensor density gets large enough. Our specific goal was thus to measure this limit, assess how many sensors are needed to reach it, and examine the effect of key parameters such as sensor type, sensor-to-brain distance, or signal-to-noise ratio (SNR). We further used simulations to investigate the opposite regime of low sensor density where the asymptotic theory breaks down.

\section{Theory}\label{Theory}

We consider a multi-channel MEG system composed of sensors surrounding the head as illustrated in Fig.~1. In this section, we present the results of a theoretical analysis of MEG spatial resolution in the asymptotic limit where a large number of sensors are distributed homogeneously on a hemispherical array (shown red in Fig.~1). This allows us to describe explicitly a measure of spatial resolution as a function of sensor type, array-to-brain distance, and SNR. The detailed developments leading to these results are relegated to two appendices; \ref{appendix-bckgd} gathers useful background and minor results on the machinery of MEG multipolar expansions, and \ref{appendix-proofs} develops our original asymptotic analysis of MEG spatial resolution. The theory is supplemented with experimental data and numerical simulations in Sections \ref{Methods} and \ref{Results}, where we also explore the regime of low sensor density outside the domain of validity of the asymptotic theory. Further intuition and practical conclusions are discussed in Section \ref{Discussion}.

%
\begin{figure}\centering
\includegraphics{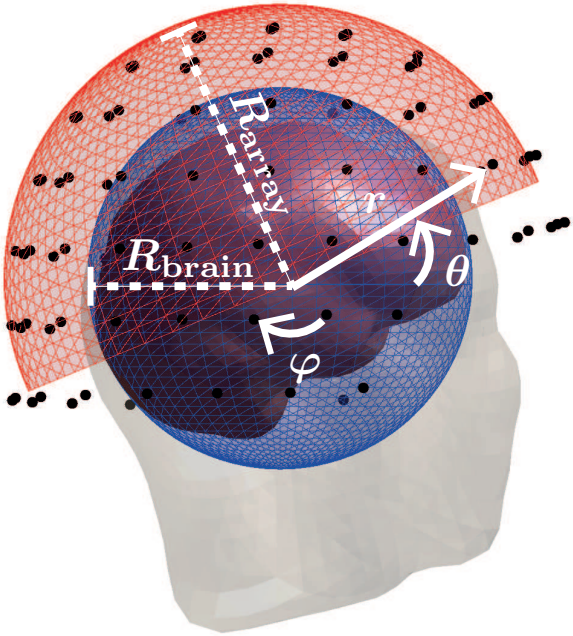}
\caption{\label{fig1} \emph{Geometric arrangement.} This illustration shows a subject's head inside a hemispherical array of radius $R_\textrm{array}$ centered on their brain (red), along with the $N=102$ sensor locations of the Neuromag MEG sensor array (black dots). The smallest concentric sphere enclosing the brain (blue) defines the anatomical brain radius $R_\textrm{brain}$. The spherical coordinate system $(r,\theta,\varphi)$ used in this paper is also indicated.}
\end{figure}

We start by describing the general framework of the theory and formulate explicitly its assumptions.

\subsection{Multipolar expansion of multi-channel MEG signals}

\paragraph{Observation model} For our purposes, a MEG array consists of a number $N$ of sensor locations where one or several components of the magnetic field or its gradient are measured. We assume that these locations can be parameterised by the angular position $\Omega$ in a suitable spherical coordinate frame centered on the subject's brain (e.g., the polar angle $\theta$ and the azimuthal angle $\varphi$ shown in Fig.~\ref{fig1}). We focus mainly on \vince{spherically shaped} arrays where the radial coordinate $r=R_\textrm{array}$ is constant (Fig.~1), but angle-dependent \vince{radial coordinates} will be allowed when we consider a realistic MEG geometry (see Sections \ref{Methods} and \ref{Results}).

Sensor measurements $\boldsymbol b(\Omega)$ may be related to point field values $\boldsymbol \phi(\Omega)$ (i.e., components of the magnetic field or its gradient at the center of the sensor) and intrinsic sensor noise $\boldsymbol \varepsilon(\Omega)$ via the observation model
\begin{equation}\label{obsmodel} \boldsymbol b(\Omega) = \boldsymbol \phi(\Omega) + \boldsymbol \varepsilon(\Omega)\, , \end{equation}
which typically holds to a good approximation in MEG systems. We allow for multimodal setups where a number $M$ of different sensors sit at the same place, so all symbols in Eq.~\eqref{obsmodel} represent $M$-vectors. For example, $M=1$ for CTF systems consisting of $N=275$ axial gradiometers and for OPM arrays composed of single-axis radial magnetometers; $M=3$ for Neuromag systems consisting of $N=102$ chipsets (Fig.~1, black dots) of one radial magnetometer and two planar gradiometers \citep[see, e.g.,][]{HariPuce} and for arrays of tri-axis OPMs \citep{Brookes2021}.

\paragraph{Spatial whiteness of intrinsic sensor noise}
We also assume that sensor noise is homogeneous and uncorrelated, so its covariance across all array locations (i.e., gathering the $N$ vectors $\boldsymbol \varepsilon(\Omega)$ in a single $NM$-vector) takes the form
\begin{equation}\label{noisecov} \mathrm{cov}(\boldsymbol \varepsilon) = \sigma^2_{\boldsymbol \varepsilon}\, \boldsymbol I\, , \end{equation}
with $\boldsymbol I$ the $NM\times NM$ identity matrix. Equation \eqref{noisecov} should be amended in multimodal setups that mix magnetometers and gradiometers (since their noise levels $\sigma_{\boldsymbol \varepsilon}$ do not carry the same physical units), but the ensuing changes are not essential so we keep it unmodified for notational simplicity.

\paragraph{Multipolar expansion} Since neuromagnetic activity is probed outside the head and works in a quasi-static regime, extra-cranial field values $\boldsymbol \phi(\Omega)$ may be subjected to an interior multipolar expansion of the form \citep{Taulu2004,Taulu2005,Tierney2022}
\begin{equation}\label{MPE-general} \boldsymbol \phi(\Omega) = \sum_{\ell,m} a_{\ell,m}\, \boldsymbol S(\Omega | \ell,m)\, .\end{equation}
The $M$-vectors $\boldsymbol S(\Omega | \ell,m)$ denote the vectorial spherical harmonics \citep{Hill} indexed by integers $\ell\geq 0$ and $-\ell\leq m\leq \ell$. See \ref{appendix-bckgd-MPEreview} (Tables \ref{TableA1} and \ref{TableA2}) for detailed expressions in cases of interest. The series \eqref{MPE-general} corresponds to a spectral decomposition of the neuromagnetic topographies in terms of angular frequency $k=\sqrt{\ell(\ell+1)}/R_\textrm{array}$, so $\ell$ indexes angular frequency \citep{Jackson,Zangwill2012}. We use in this work the inverse of the radial coordinate $r$ relative to the brain sphere radius $R_\textrm{brain}$ (see Fig.~\ref{fig1}) as expansion parameter (\ref{appendix-bckgd-MPEreview}). In this way, all multipole moment coefficients $a_{\ell,m}$ share the same physical unit $[\mathrm{T}\cdot \mathrm{m}]$ and may be compared numerically. This allows to formulate the following hypothesis that is fundamental to our analysis of MEG spatial resolution.

\paragraph{Maximum-entropy hypothesis} We assume that all multipole moments are uncorrelated and of equal variance, i.e.,
\begin{equation}\label{maxentropy} \mathrm{cov}(\boldsymbol a) = \sigma^2_{\boldsymbol a}\, \boldsymbol I \end{equation}
using formal notations where the coefficients $a_{\ell,m}$ are gathered into an infinite column vector $\boldsymbol a$ and where $\boldsymbol I$ denotes an infinite square identity matrix. This corresponds to a situation of ``maximum entropy'' where brain activity is spatially unstructured and involves all spatial scales equally, from microscopic (e.g., single-channel synaptic currents) to macroscopic (i.e., whole-brain network) levels. That is both unphysical and biologically unrealistic, but nevertheless useful for exploring the limits of MEG spatial resolution. Extra-cranial measurements are at best sensitive to the mean activity of neural populations, but the assumption \eqref{maxentropy} also includes undetectable microscopic and other non-physiological electrical source configurations, leading to an overestimation of MEG spatial resolution. This overestimation is illustrated with experimental data in Section \ref{Results}.

A solution to this important caveat is to abandon a direct physiological interpretation of the two parameters of the MEG multipolar expansion model, i.e., the brain sphere radius $R_\textrm{brain}$ and the multipole amplitude $\sigma_{\boldsymbol a}$. Instead, we propose to treat them as effective parameters of the theory to be assessed empirically from data. According to the hypothesis \eqref{maxentropy}, a brain sphere with radius $R_\textrm{brain}$ estimated na\"ively from anatomy (blue sphere in Fig.~\ref{fig1}) would include highly localized neural activity right under (or even slightly above) the brain convexity beneath the scalp. Such configuration must be associated with a focal field topography and thus high MEG spatial resolution, but it might not be representative of the experimental data at hand. In turn, this might lead to an underestimation of the multipole amplitude parameter $\sigma_{\boldsymbol a}$, which can be determined from the SNR estimate
\begin{equation}\label{snr-def} \textrm{SNR} = \tfrac{1}{NM} \mathrm{Tr}\left[  \mathrm{cov}(\boldsymbol \varepsilon)^{-1} \mathrm{cov}(\boldsymbol b)\right] \end{equation}
of MEG recordings \eqref{obsmodel} via the relation (\ref{appendix-bckgd-MPEminor})
\begin{equation}\label{sigasigeps-est} \frac{\sigma_{\boldsymbol a}^2}{\sigma_{\boldsymbol \varepsilon}^2}=\frac{NM\, (\textrm{SNR}-1)}{\mathrm{Tr}( \boldsymbol S\, \boldsymbol S^\dagger )}\, \cdot\end{equation}
Here, $\boldsymbol S$ is a formal matrix with an infinite number of columns indexed by $(\ell,m)$, each column gathering the $NM$ elements of the $M$-vectors $\boldsymbol S(\Omega | \ell,m)$ at the $N$ sensor locations of the MEG array. We describe in Section \ref{Methods} how to combine anatomical brain images and MEG recordings in order to determine functional estimates of $R_\textrm{brain}$ and $\sigma_{\boldsymbol a}/\sigma_{\boldsymbol \varepsilon}$ and obtain physiologically meaningful MEG multipolar expansions.

\paragraph{Spatial resolution from multipolar expansions} The vectorial spherical harmonics $\boldsymbol S(\Omega | \ell,m)$ in Eq.~\eqref{MPE-general} measure the sensitivity of MEG sensors to neuromagnetic fields with definite angular frequency $\ell$. Sensitivity decreases exponentially fast for highly focal neuromagnetic topographies characterized by large values of $\ell$ 
(\ref{appendix-bckgd-MPEreview}). This exponential suppression embodies the physical smoothing process that neuromagnetic fields undergo as they propagate from brain sources to sensors. On the other hand, intrinsic sensor noise contributes equally at all the spatial scales sampled by the MEG array; this is embodied by the spatial whiteness assumption \eqref{noisecov}. It is the interaction of these two features that inherently limits the sensitivity of MEG data to focal brain activity. Measurement noise is typically subdominant at low angular frequency but overshadows focal neuromagnetic activity at high angular frequency. Effectively, noise should cut off the expansion \eqref{MPE-general} at a critical value $\ell=\ell_*$ where this cross-over occurs. This idea is the basis of signal-space separation \citep{Taulu2004, Taulu2005}. We leverage it here and seek to measure MEG spatial resolution using the critical value $\ell_*$, since it corresponds to the smallest spatial scale that is experimentally accessible.

Our main goal is to determine explicitly this spatial resolution index $\ell_*$. Quite amazingly, this turns out possible for \vince{hemispherically shaped} MEG arrays in the limit $N\rightarrow\infty$ corresponding to an infinitely dense, homogeneous sensor coverage (Fig.~\ref{fig1}). The usefulness of considering this situation inaccessible to both experiment and simulations is that it allows precisely to assess the limits of MEG spatial resolution and how they depend on sensor type, array-to-brain distance, and SNR.

\paragraph{Signal-space dimension} In situations where the validity of the asymptotic theory is not settled, we will resort to signal-space dimension as proxy measure of spatial resolution. We define it here as the number $\nu$ of degrees of freedom contained in brain MEG signals and estimated according to
\begin{equation} \label{sdof-a} \nu = \#\left\{\, \textrm{eigenvalues $\lambda_u^2$ of $\boldsymbol S\, \boldsymbol S^\dagger$ with } \lambda_u^2 >  \sigma_{\boldsymbol \varepsilon}^2/\sigma^2_{\boldsymbol a} \, \right\} \, .
\end{equation}
This corresponds to the number of linearly independent neuromagnetic topographies (i.e., eigenvectors of the $NM\times NM$ matrix $\boldsymbol S\, \boldsymbol S^\dagger$) whose contribution (measured by their eigenvalue $\lambda_u^2$) exceeds noise level ($\sigma_{\boldsymbol \varepsilon}^2/\sigma^2_{\boldsymbol a}$) and thus is experimentally detectable (\ref{appendix-bckgd-MPEminor}). These topographies span what is known as the signal space \citep{Taulu2004, Taulu2005}.

\vince{The s}ignal-space dimension $\nu$ assesses the information content of MEG data rather than their spatial resolution \emph{per se}. It must be commensurate to spatial resolution since access to more focal details should increase the number of detectable topographies. Yet, like any other complexity metric (may they be linear dimensions or non-linear, information-theoretic capacities), it turns out to mix spatial resolution and other geometric factors of the MEG array. This is demonstrated below.

\subsection{Asymptotic regime of high-density MEG}\label{largeNanalysis}

\paragraph{Spatial resolution index in the large-$N$ limit} We present here our main theoretical result about asymptotically high-density MEG arrays with hemispherical geometry and homogeneously distributed sensors (Fig.~\ref{fig1}). Mathematical analysis of the limit $N\rightarrow\infty$ developed from \ref{appendix-proofs-spectrum} to \ref{appendix-proofs-largeNsoln} determines the spatial resolution index as 
\begin{equation}\label{lstar-subleading} \ell_* = \frac{\log \left[ \frac{N}{4\pi\, r^{2+\mathrm{deg}(\mathsf P)}} \frac{\sigma_{\boldsymbol a}^2}{\sigma_{\boldsymbol \varepsilon}^2}\, \mathsf P\left(\frac{\log N}{2\log r}\right) \right]}{2 \log r} + \mathcal O\left( \frac{\log\log N}{\log N}\right)\, . \end{equation}
This result enables the quantitative measurement of MEG spatial resolution as a function of the number $N$ of sensors, the sensor-to-brain distance $r=R_\textrm{array}/R_\textrm{brain}$ (i.e., the radius of the hemispherical array relative to that of the brain sphere)
, and the multipole SNR parameter $\sigma_{\boldsymbol a}/\sigma_{\boldsymbol \varepsilon}$. It also depends on the type of sensors composing the MEG array through a polynomial $\mathsf P$ that is identified in \ref{appendix-proofs-spectrum} (Table \ref{TableA3}).

\paragraph{Properties of high-density MEG spatial resolution}
Let us describe the effect of different MEG array characteristics disclosed by Eq.~\eqref{lstar-subleading}. See \ref{appendix-proofs-properties} for more details.

\begin{enumerate}[(i)]
\item\label{item-density} \emph{Sensor density.} The spatial resolution index $\ell_*$ exhibits a logarithmic divergence as $N$ grows indefinitely. In other words, \emph{the limits of spatial resolution increase without bound (albeit very slowly) as MEG arrays become denser}. This observation contradicts our initial expectation that spatial resolution would converge towards a definite limit controlled by magnetic field smoothness. Rather, it is the extreme slowness of this divergence that corresponds to the constraints imposed by field smoothness.

\item\label{item-radius} \emph{Sensor-to-brain distance.} The rate at which the spatial resolution diverges turns out to be controlled by the parameter $r$ and not by any other MEG characteristics. The dependence in other characteristics is milder because both sensor type and SNR only contribute through sub-leading corrections that are small compared to the leading divergence. This means that \emph{the limits of spatial resolution are mostly modulated by the sensor-to-brain distance}. In fact, $\ell_*$ appears to increase without bound as the sensor array approaches the brain surface ($r\rightarrow 1$), i.e., spatial resolution improves drastically as the MEG array approaches the brain. Nevertheless, this divergence is, in a sense, only an artifact as the asymptotic theory breaks down before sensors reach the brain (see Section \ref{Results}).

\item\label{item-type} \emph{Sensor type.} The magnetometric or gradiometric nature of MEG sensors makes a sub-leading contribution to $\ell_*$ that is nevertheless numerically significant, because it also exhibits a divergence as $N$ grows indefinitely (although an even slower one). It turns out that this contribution is twice larger for gradiometers, so we conclude that \emph{gradiometric arrays exhibit moderately higher limits of spatial resolution than magnetometric arrays (at similarly large number $N$ of sensors)}. On the other hand, \emph{the number of recorded components or their orientation only have a minute impact} as their contribution is either finite or negligibly small at large $N$.

\item\label{item-snr} \emph{Multipole SNR.} Likewise, \emph{the SNR makes a subtle, finite contribution} that is negligible. 

\end{enumerate}

\paragraph{Signal-space dimension in the large-$N$ limit}
We further demonstrate in \ref{appendix-proofs-spectrum} that the signal-space dimension $\nu$ may be expressed in terms of the spatial resolution index $\ell_*$ by merely counting the number of vectorial spherical harmonics whose contribution to MEG signals exceeds noise level, i.e., for which $\ell\leq \ell_*$. A straightforward count \cite[as done in, e.g., ][]{Taulu2005,Tierney2022} suggests a value
\begin{equation}\label{nu-sphere} \nu_\mathbb S = (\ell_*+1)^2 \end{equation}
but that is not quite right. In fact, this relation is only valid for a hypothetical MEG array that covers a complete sphere $\mathbb S$ enclosing the brain (notwithstanding that this would be nonsensical from the experimental standpoint); this is emphasized by the subscript attached to the symbol $\nu$ in Eq.~\eqref{nu-sphere}. A proper analysis of MEG multipolar expansions on a hemisphere $\mathbb H$ (\ref{appendix-bckgd-MPEminorH}) reveals instead that
\begin{equation}\label{nu-hemisphere} \nu_\mathbb{H}=\tfrac{1}{2} (\ell_*+1)(\ell_*+2)\, . \end{equation}
This is approximately twice smaller, which reflects the halving of sensor coverage compared to the whole sphere. The dependence in sensor coverage demonstrates the difference between the spatial resolution index (which is the same for spherical and hemispherical MEG; see \ref{appendix-proofs-spectrum}) and complexity metrics such as signal-space dimension (see also \ref{appendix-proofs-properties}).

\section{Methods}\label{Methods}

We describe numerical and experimental methods to estimate the parameters of MEG multipolar expansions ($r$ and $\sigma_{\boldsymbol a}/\sigma_{\boldsymbol \varepsilon}$), examine the domain of validity for our asymptotic theory (i.e., how large the number $N$ of sensors must be to ensure the quantitative accuracy of Eq.~\ref{lstar-subleading}), explore what happens at low sensor density outside this domain of validity, and finally measure quantitatively the limits of MEG spatial resolution.

\subsection{Numerical evaluation of signal-space dimension}\label{nu-numerics}

At the core of our numerical experiments is an estimation of signal-space dimension that works whatever the MEG array (e.g., hemispherical or Neuromag geometries illustrated in Fig.~\ref{fig1}) and whatever the number $N$ of sensors (as long as it is not so large that simulations become untractable).

The $NM\times NM$ matrix $\boldsymbol S\, \boldsymbol S^\dagger$ appearing in Eq.~\eqref{sdof-a} gathers $M\times M$ blocks of the form $ \sum_{\ell=0}^\infty \sum_{m=-\ell}^\ell \boldsymbol S(\Omega|\ell,m)\, \boldsymbol S(\Omega'|\ell,m)^\dagger$, where $\Omega, \Omega'$ run over the $N$ sensor locations. The infinite sum over $\ell$ was evaluated by computing terms at successive values $\ell=0,1,2,\ldots$ and adding them iteratively until numerical convergence (which is guaranteed). The vectorial spherical harmonics $\boldsymbol S(\Omega|\ell,m)$ were evaluated at angular ($\theta,\varphi$) and radial ($r$) coordinates corresponding to sensor locations of the MEG arrays described below and for different sensor types (radial and tri-axis magnetometers, axial and planar gradiometers, see \ref{appendix-bckgd-MPEreview}). Summation was performed over the first hundred terms ($0\leq \ell\leq 100$) and then continued iteratively until the last term to add got small enough; as precise criterion, we required that the squared Frobenius norm of the current, $\ell^\textrm{th}$ term
, relative to that of the partial sum over all $\ell-1$ previous terms, reach below $10^{-5}$.
The signal-space dimension \eqref{sdof-a} was then evaluated by diagonalizing the partial sum and counting the number of eigenvalues above threshold $\sigma_{\boldsymbol \varepsilon}^2/\sigma_{\boldsymbol a}^2$.

\subsection{Anatomical MEG expansion parameters}\label{param-anat}

We considered MEG resting-state data of 14 healthy adult subjects used in previous studies \citep{Coquelet2020,Coquelet2022}, to which we refer for details. Briefly, MEG signals were acquired at rest (5 min, 0.1--330 Hz analog bandpass, 1 kHz sampling rate) using a Neuromag Vectorview system (MEGIN Oy, Helsinki, Finland) and denoised using signal-space separation \citep[Maxfilter v2.2 with default parameters $\ell_\textrm{in}=8$ and $\ell_\textrm{out}=3$, MEGIN; ][]{Taulu2004,Taulu2005} and independent component analysis \citep{HyvarinenOja2000}. We used these data to extract geometric information needed to construct the $\boldsymbol S$ matrices (\ref{appendix-bckgd-MPEreview}) and functional information related to the SNR.

The Neuromag MEG array is composed of sensors located at $N=102$ locations (Fig.~\ref{fig1}) comprising one radial magnetometer and two orthogonal planar gradiometers. We used individual brain magnetic resonance images (MRIs) co-registered with the MEG array to define subject-specific spherical coordinates of each sensor location. The coordinate origin was set at the centre of the sphere fitted to the vertices of the scalp surface obtained after tissue segmentation \citep[Freesurfer, Martinos Center for Biomedical Imaging, Massachussetts, USA; ][]{Fischl2012}. This allowed to assign radial (distance from origin) and  angular coordinates to each sensor. The array radius $R_\textrm{array}$ was defined as the root-mean-square of all 102 radial coordinates, and the anatomical brain sphere of radius $R_\textrm{brain}$ was determined as the smallest sphere centered on the origin that encloses the inner skull surface (Fig.~\ref{fig1}, blue). The ratio $R_\textrm{array}/R_\textrm{brain}$ determined the ``anatomical'' estimate of the expansion parameter $r$ for the Neuromag MEG array. 
To illustrate the impact of array-to-brain distance, we also considered a virtual OPM array placed 6.5 mm above scalp and thereby obtained an anatomical estimate of $r$ corresponding to scalp MEG. The 6.5-mm height corresponds to the center location of the alkali vapour cell in Gen-2 QZFM sensors (QuSpin Inc., Colorado, USA) placed directly on scalp.

The SNR associated with Neuromag MEG recordings at rest was estimated for magnetometers ($N=102$, $M=1$) and planar gradiometers ($N=102$, $M=2$) separately according to Eq.~\eqref{snr-def}, with the $NM\times NM$ data covariance $\mathrm{cov}(\boldsymbol b)$ extracted from the resting-state recordings and the noise covariance $\mathrm{cov}(\boldsymbol \varepsilon)$, from empty-room recordings. The noise covariance was regularized prior to inversion by adding 10\% of the mean sensor variance to its diagonal. Combining this SNR measure with the computation of the corresponding $\boldsymbol S\, \boldsymbol S^\dagger$ matrix (based on the above geometric information and on Section \ref{nu-numerics}) and with Eq.~\eqref{sigasigeps-est}, we could then estimate the multipole SNR parameter $\sigma_{\boldsymbol a}/\sigma_{\boldsymbol \varepsilon}$.

The MEG multipolar expansion models constructed in this way will be referred to as ``anatomical MEG'' as they are inferred from the actual brain size of subjects.

\subsection{Functional MEG expansion parameters}\label{param-func}

We also determined ``functional MEG'' multipolar expansion models, in which the relative radius $r$ and multipole SNR $\sigma_{\boldsymbol a}/\sigma_{\boldsymbol \varepsilon}$ are estimated by fitting the measure \eqref{sdof-a} of the signal-space dimension $\nu$ to another measure $\nu_\textsc{fwd}$ based on MEG forward modeling, i.e., explicit simulations of neuromagnetic field propagation from brain sources to sensors. The rationale is that, although MEG forward models cannot be used to probe the large-$N$ limit, they describe finite-$N$ MEG measurements under \vince{biologically realistic} conditions and only contain explicit parameters (contrary to the multipolar expansion, as explained below Eq.~\ref{maxentropy}).

The approach based on MEG forward modeling is formally similar to that based on multipolar expansions and Eq.~\eqref{sdof-a}; the only difference is that the $\boldsymbol S$ matrix is replaced by the leadfield matrix $\boldsymbol L$ and the set of multipole moments $\boldsymbol a$, by the distribution $\boldsymbol j$ of electrical currents over the source space. The setup is actually that of linear MEG source projection, i.e., field measurements \eqref{MPE-general} are expressed as $\boldsymbol \phi = \boldsymbol L \, \boldsymbol j$ and the source covariance is assumed diagonal, $\mathrm{cov}(\boldsymbol j)=\sigma_{\boldsymbol j}^2\, \boldsymbol I$ (which is a version of the ``maximum-entropy'' condition; compare with Eq.~\ref{maxentropy}). Source variance $\sigma_{\boldsymbol j}^2$ was inferred from an analog of Eq.~\eqref{sigasigeps-est} with $\boldsymbol S\, \boldsymbol S^\dagger$ replaced by the leadfield covariance $\boldsymbol L\, \boldsymbol L^\mathrm{T}$. The number $\nu_\textsc{fwd}$ of spatial degrees of freedom corresponding to the MEG forward model was then obtained by counting the number of eigenvalues of $\boldsymbol L\, \boldsymbol L^\mathrm{T}$ exceeding $\sigma_{\boldsymbol \varepsilon}^2/\sigma_{\boldsymbol j}^2$, in complete analogy with Eq.~\eqref{sdof-a}.
In practice, we computed individual MEG forward models using MRI tissue segmentation and the three-layer boundary element method implemented in MNE-C \citep{Gramfort2014}. The brain volume was discretized into a regular 5-mm cubic lattice on which three orthogonal unit current dipoles were placed. Sensor locations corresponded to the Neuromag array (cryogenic MEG) or the virtual OPMs (scalp MEG) co-registered to the MRI. The resulting leadfields allowed us to generate MEG resting-state estimates $\nu_\textsc{fwd}$ of the signal-space dimension.

We then determined the parameters $r=R_\textrm{array}/R_\textrm{brain}$ and $\sigma_{\boldsymbol a}/\sigma_{\boldsymbol \varepsilon}$ for which the signal-space dimension \eqref{sdof-a} coincides with the MEG forward model estimate, i.e., $\nu=\nu_\textsc{fwd}$. We solved this problem in a pragmatic way with an iterative two-step optimization algorithm that controls for the mutual influence of these two parameters. In a first step, $\nu$ was computed numerically (Section \ref{nu-numerics}) over a predefined grid of brain radii $R_\textrm{brain}$ (from 1 mm to 10 cm with 1-mm spacing), using the value of $\sigma_{\boldsymbol a}/\sigma_{\boldsymbol \varepsilon}$ determined at the previous iteration. The radius that best fits $\nu$ to $\nu_\textsc{fwd}$ was then selected. In the second step, the multipole SNR parameter was updated using Eq.~\eqref{sigasigeps-est}. The initial condition was taken as the anatomical estimate and the algorithm was stopped once the fit error $\nu-\nu_\textsc{fwd}$ reached below $10^{-3}$. Although we had no guarantee of convergence, this procedure generated definite functional parameters separately for magnetometers and planar gradiometers.

\subsection{Simulated hemispherical MEG arrays}\label{sph-numerics}

We simulated hemispherical MEG arrays with a variable number of sensors ranging from small ($N=5$) to fairly large ($N=2500$). For each value of $N$, a homogeneous grid of sensor locations covering a hemisphere was defined by splitting the north hemisphere parameterized by $0\leq \theta\leq \pi/2$ into a number $n$ of \vince{equally spaced} circles of latitude, and placing a latitude-dependent number $m$ of \vince{equally spaced} sensors on each circle. The pole $\theta=0$ was excluded to enable efficient algebraic computation of spherical harmonics derivatives (\ref{appendix-bckgd-MPEreviewYlm}). We set $n$ to the integer nearest to $\sqrt{N}$ and $m$ to the integer nearest to $n\pi \sin\theta / 2$; this ensured that the solid angle per sensor was approximately constant and equal to $2\pi/N$, i.e., sensor coverage was homogeneous.

We ran the numerical computation of signal-space dimension as a function of $N$ for hemispherical MEG arrays composed of a number $N$ of sensors, with parameters set either to the anatomical or functional multipolar expansions of cryogenic or scalp MEG.

\section{Results}\label{Results}

\subsection{Parameters for MEG multipolar expansions}\label{res-param}

\begin{figure*}\centering
\includegraphics[scale=0.75]{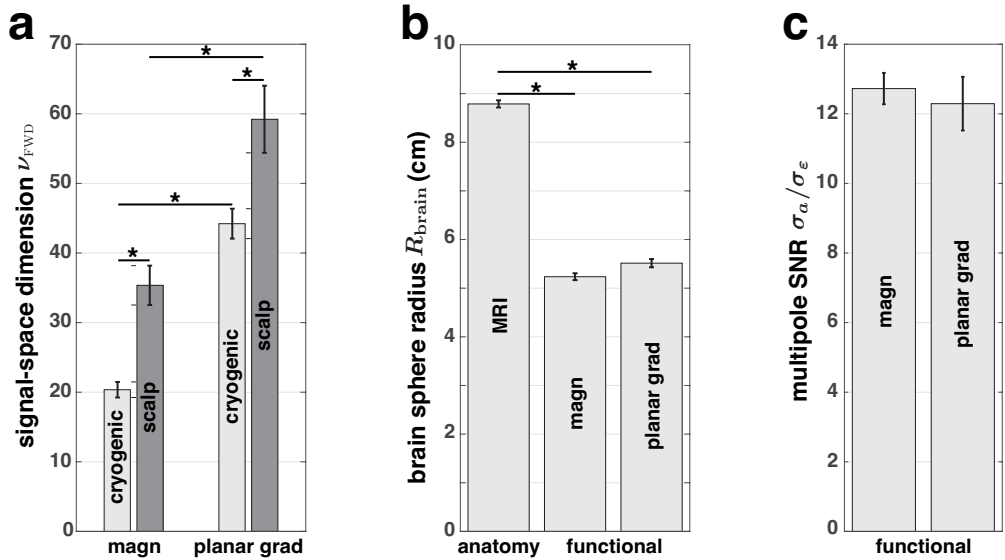}
\caption{\label{fig2}\emph{Model parameters for anatomical and functional MEG multipolar expansions.} \textbf{a.} Signal-space dimension $\nu_\textsc{fwd}$ based on forward models corresponding to cryogenic MEG ($N=102$ SQUIDs, Neuromag system, light grey) and scalp MEG ($N=102$ virtual OPMs, dark grey). Magnetometers and planar gradiometers are considered separately. \textbf{b.} Brain sphere radius $R_\textrm{brain}$ estimated from anatomy (MRI) or functional data (magnetometers and planar gradiometers separately). \textbf{c.} Multipole SNR parameter $\sigma_{\boldsymbol a}/\sigma_{\boldsymbol \varepsilon}$ estimated from functional data. Functional estimates correspond to cryogenic MEG. Plots report mean $\pm$ SEM over subjects. magn: magnetometer, grad: gradiometer, MRI: magnetic resonance imaging, SNR: signal-to-noise ratio, SEM: standard error of the mean, $\star$: $p<0.05$ (ANOVA, post-hoc tests).}
\end{figure*}

We started our numerical exploration of MEG spatial resolution with the generic notion of signal-space dimension \eqref{sdof-a}.
Figure \ref{fig2}a reports experimental estimates 
obtained from the $N=102$ magnetometers or the $N=102$ pairs of planar gradiometers in Neuromag resting-state recordings (cryogenic MEG; see Fig.~\ref{fig1}) and from corresponding virtual OPM arrays (scalp MEG). Signal-space dimension was larger for gradiometers than magnetometers, and for scalp than cryogenic MEG (two-way ANOVA, main effects at $p<10^{-5}$, no significant interaction). Given that signal-space dimension reflects both spatial resolution and sensor coverage (see Eqs.~\ref{nu-sphere} and \ref{nu-hemisphere}), and that sensor coverage is the same in the four cases compared in Fig.~\ref{fig2}a, these effects illustrate indirectly the impact of sensor type and sensor-to-brain distance on MEG spatial resolution. 

%

We then selected physiologically relevant parameters for MEG multipolar expansion models by combining anatomical MRIs and the cryogenic MEG data. Figure \ref{fig2}b compares the brain sphere radius inferred from anatomy (anatomical MEG; Fig.~\ref{fig1}, blue) and functional estimates designed to faithfully reproduce the experimental values of cryogenic MEG signal-space dimension (functional MEG). We observed a difference between the anatomical and the functional brain spheres (one-way ANOVA, $p=3.8\times 10^{-7}$), the latter being smaller than the former (post-hoc $p<4.0\times 10^{-4}$). The surprisingly small brain radii of functional MEG ($5.3$--$5.5$ cm) compared to anatomy ($8.8$ cm) suggests to interpret these functional brain spheres as averages of concentric spherical layers within the anatomical brain probing the sources of MEG resting-state activity, from neocortical to deep cortical regions (see Section \ref{Discussion} for further discussion). The functional radii obtained from magnetometers and gradiometers were similar (post-hoc $p=0.31$), so in subsequent analyses they were averaged to generate a single functional brain radius. Combining these two estimates with the geometry of either cryogenic or scalp MEG yielded four distinct values of interest for the sensor-to-brain distance parameter (Table \ref{Table4}).

\begin{table}[ht]
\begin{center}
\begin{tabu}{|[1pt]c|[1pt]c|[1pt]c|[1pt]}
    \tabucline[1pt]{1-3}
    \textbf{parameter} & \textbf{estimate} & \textbf{mean $\pm$ SD} \\
    \tabucline[1pt]{1-3}
    \multirow{2}{*}{MEG array size ($R_\textrm{array}$)} & cryo & $12.2\pm0.2$ cm \\ [-1pt] \tabucline[0.5pt]{2-3} & scalp & $9.6\pm0.3$ cm \\
    \tabucline[1pt]{1-3}
    \multirow{2}{*}{Brain sphere radius ($R_\textrm{brain}$)} & anat & $8.8\pm0.3$ cm \\ [-1pt] \tabucline[0.5pt]{2-3} & func & $5.4\pm0.3$ cm \\
    \tabucline[1pt]{1-3}
    \multirow{4}{*}{Sensor-to-brain radius ($r$)} & func-cryo\textsuperscript{(a)} & $2.28\pm0.10$ \\ [-1pt] \tabucline[0.5pt]{2-3} & func-scalp\textsuperscript{(b)} & $1.79\pm0.05$ \\ [-1pt] \tabucline[0.5pt]{2-3} & anat-cryo\textsuperscript{(c)} & $1.39\pm0.03$ \\ [-1pt] \tabucline[0.5pt]{2-3} & anat-scalp\textsuperscript{(d)} & $1.10\pm0.02$ \\
    \tabucline[1pt]{1-3}
    Multipole SNR ($\sigma_{\boldsymbol a}/\sigma_{\boldsymbol \varepsilon}$) & func & $12.5\pm 1.7$ \\
    \tabucline[1pt]{1-3}
    \end{tabu}
\end{center}
\caption{\label{Table4}{\it Model parameters for anatomical and functional MEG multipolar expansions.} The sensor-to-brain radius $r$ corresponds to the ratio $R_\textrm{array}/R_\textrm{brain}$. The group means of the model parameters $r$ and $\sigma_{\boldsymbol a}/\sigma_{\boldsymbol \varepsilon}$ were used in quantitative applications of the theory. Labels (a)--(d) are introduced for reference in text and figures. SNR: signal-to-noise ratio, cryo: cryogenic MEG (Neuromag system), scalp: scalp MEG (virtual OPMs), anat: anatomical estimate, func: functional estimate (averaged over magnetometric and gradiometric estimates), SD: standard deviation.}
\end{table}

In the same vein, Figure \ref{fig2}c shows that magnetometers and gradiometers led to similar functional estimates of the multipole SNR parameter $\sigma_{\boldsymbol a}/\sigma_{\boldsymbol \varepsilon}$ (one-way ANOVA, $p=0.61$), so we averaged their values as well (Table \ref{Table4}).

\begin{figure}\centering
\includegraphics[scale=0.8]{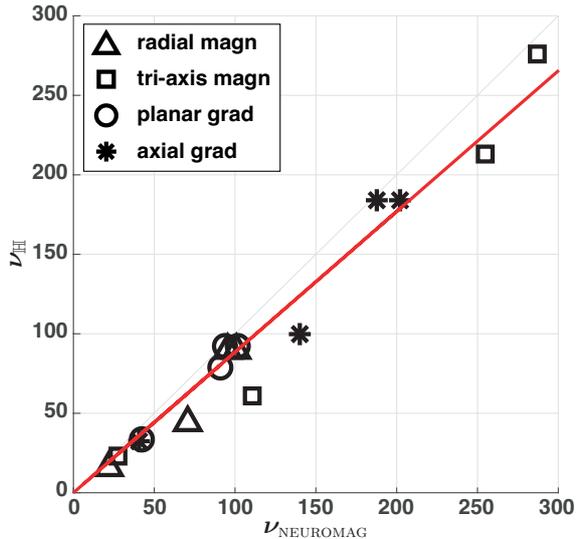}
\caption{\label{figR1} \emph{Dimensionality of Neuromag-like MEG and hemispherical MEG.} This plot compares two estimates of the signal-space dimension, one where $N=102$ sensors are arranged according to a realistic MEG geometry ($\nu_\textsc{neuromag}$; Neuromag SQUIDs or corresponding virtual scalp OPMs) and the other where $N=102$ sensors cover homogeneously a hemisphere of same array radius ($\nu_\mathbb{H}$). Each value reports the group average obtained from simulating different sensor types (radial or tri-axis magnetometers, planar or axial gradiometers) with the anatomical or functional parameter estimates listed in Table \ref{Table4}. The thick line superimposed to data points shows the corresponding linear regression model $\nu_\mathbb{H}=a \, \nu_\textsc{neuromag}$ with slope $a=0.89$ ($95\%$ confidence interval, $0.83 < a < 0.94$). magn: magnetometer, grad: gradiometer.}
\end{figure}

We used the resulting parameter estimates in subsequent analyses to construct multipolar expansion models of hemispherical MEG with varying number of sensors and various sensor types. These estimates appeared fortunately insensitive to sensor type, but  still might depend on the precise shape of the MEG array. Figure \ref{figR1} assesses the impact of replacing the Neuromag-like geometry (experimental SQUID arrays or corresponding virtual OPMs) by an idealized hemisphere of same radius and same number $N=102$ of sensors. Signal-space dimension was significantly lower with hemispherical MEG in all cases ($t$ tests, $p<3\times 10^{-6}$) except for anatomical scalp MEG (labelled d in Table \ref{Table4}) where this effect was statistically marginal (magnetometers, $p=0.01$; gradiometers, $p>0.08$). The underestimation factor was $89\%\pm 6\%$ (confidence interval for linear regression slope; Fig.~\ref{figR1}, thick line), which is compatible with the coverage area of a perfect hemisphere being $87\%$ smaller than that of the Neuromag array . This suggests that this reduction in signal-space dimension merely reflects a change in sensor coverage but not in spatial resolution \emph{per se} \vince{(see Section \ref{Discussion} for further discussion)}. That is reminiscent of our theoretical comparison of hemispherical and spherical MEG (Section \ref{Theory}).

%

\subsection{Domain of validity of the large-$N$ limit}\label{res-domain}

\begin{figure*}\centering
\includegraphics[scale=0.9]{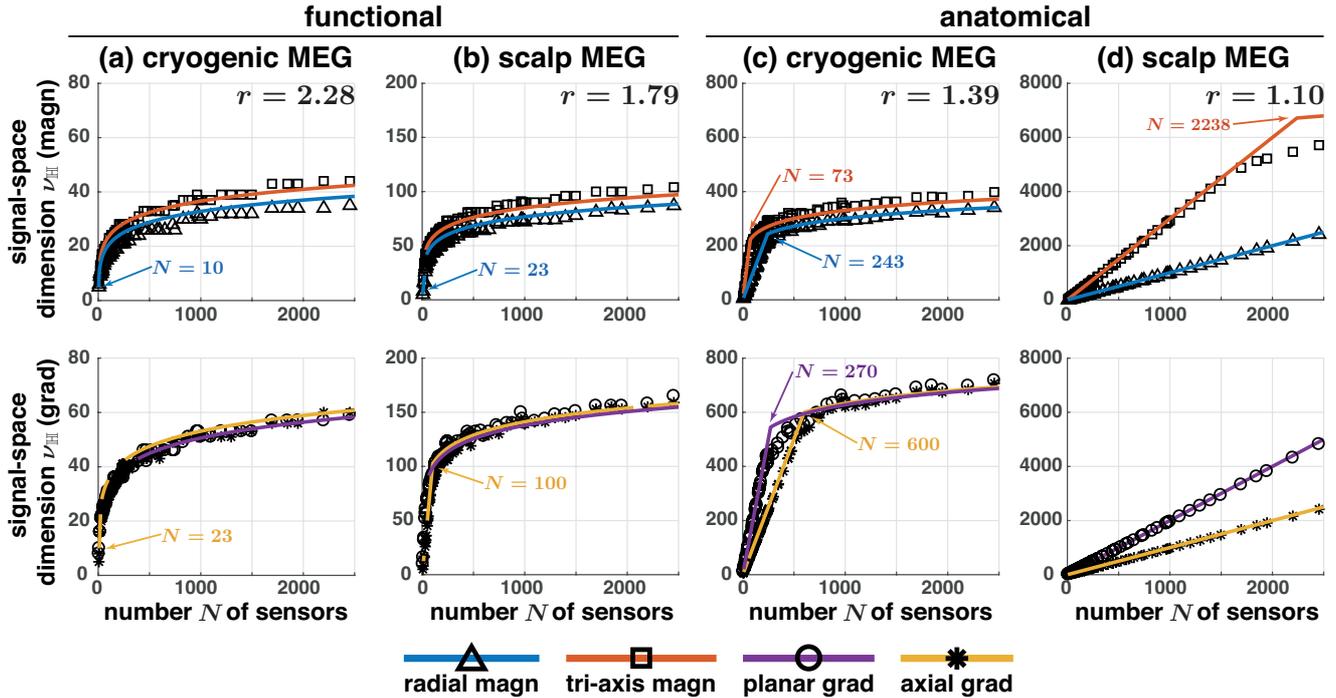}
\caption{\label{fig3} \emph{Dimensionality of hemispherical MEG multipolar expansions in the asymptotically high-density and in the low-density regimes.} Data points correspond to the signal-space dimension $\nu_{\mathbb H}$ obtained from simulations with a varying number $N$ of sensors and different sensor types (\textbf{top}: magnetometers, \textbf{bottom}: gradiometers). The sensor-to-brain radius $r$ was set according to functional (\textbf{left}) or anatomical (\textbf{right}) MEG expansion parameters corresponding to cryogenic (panels \textbf{a}, \textbf{c}) or scalp MEG (panels \textbf{b}, \textbf{d}) and the multipole SNR, to $\sigma_{\boldsymbol a}/\sigma_{\boldsymbol \varepsilon}=12.5$ (Table \ref{Table4}). Curves superimposed to the data points combine the linear regime observed at low sensor density (small $N$, Eq.~\ref{nu-linear}) and the asymptotic regime described at high-density (large-$N$ theory, Eqs.~\ref{nu-hemisphere} and \ref{lstar-subleading}). Arrows indicate the transition between the two regimes and separate their respective domains of validity. magn: magnetometer, grad: gradiometer.}
\end{figure*}

Figure \ref{fig3} reports numerical estimates of the signal-space dimension for simulated hemispherical MEG arrays from small to large numbers $N$ of sensors, once again using functional and anatomical MEG multipolar expansion models (Table \ref{Table4}). Signal-space dimension increased monotonically with a nonlinear slowing down visible when $N$ got large enough (except for anatomical scalp MEG, see Fig.~\ref{fig3}d). Because sensor coverage is identical in all simulations, this increase may be ascribed to higher spatial resolution, so adding sensors improves spatial resolution but with a smaller and smaller gain per extra sensor. This nonlinearity fit well the theoretical prediction (Eq.~\ref{nu-hemisphere} combined with Eq.~\ref{lstar-subleading}), at least for functional MEG based on resting-state recordings (see curves superimposed to Fig.~\ref{fig3}a,b). On the other hand, moving closer to the scalp revealed that a linear increase precedes the large-$N$ nonlinearity (Fig.~\ref{fig3}c,d). This was not clearly visible in functional MEG, but it was particularly obvious for anatomical scalp MEG (Fig.~\ref{fig3}d) where the nonlinear slowing down was barely reached or not at all. 

%

We used this linear regime to determine the domain of validity of the asymptotic theory. 
A careful glance at Fig.~\ref{fig3} (especially panel d) revealed that signal-space dimension $\nu_{\mathbb H}$ initially grows as $N$ for radial magnetometers and axial gradiometers, $2N$ for planar gradiometers, and $3N$ for tri-axis magnetometers. These observations may be summarized as
\begin{equation}\label{nu-linear} \nu_{\mathbb H} = MN\, \qquad \textrm{(low-density regime)}\, .  \end{equation}
%
We conclude that \emph{the number $M$ of recorded components per sensor (\ref{appendix-bckgd-MPEreview}) controls spatial resolution gains in MEG arrays with low sensor density}. In hindsight, this merely corresponds to the fact that any extra sensor component adds a new, fully independent signal as long as sensor separation is larger than the size of magnetic field smoothness. This low-density regime extended to larger values of $N$ as MEG sensors approached the scalp (compare curves from leftmost to rightmost plots of Fig.~\ref{fig3}). This reflects a reduction of field smoothness as sensors may then be closer while still bringing independent information.

We tentatively identified the transition between low- and high-density regimes as the points where the linear and large-$N$ predictions coincide (indicated by arrows on Fig.~\ref{fig3} wherever such transitions could be found). 
We combined the two regimes by gluing the linear description \eqref{nu-linear} on the left of the transition and the large-$N$ solution \eqref{lstar-subleading} on the right (curves superimposed to Fig.~\ref{fig3}). Two signs betray the somewhat artifical nature of this recombination; the non-smooth corner at the transition point and the overestimation of signal-space dimension around this point where neither regime is fully valid (see, e.g., anatomical MEG cryogenic planar gradiometers in Fig.~\ref{fig3}c or tri-axis magnetometers in Fig.~\ref{fig3}d). We will nevertheless use this hard transition to estimate the domain of validity of the asymptotic theory, and apply the linear description \eqref{nu-linear} outside of this domain.

\subsection{Quantitative limits of MEG spatial resolution}\label{res-lstar}

\begin{figure*}\centering
\includegraphics[scale=1]{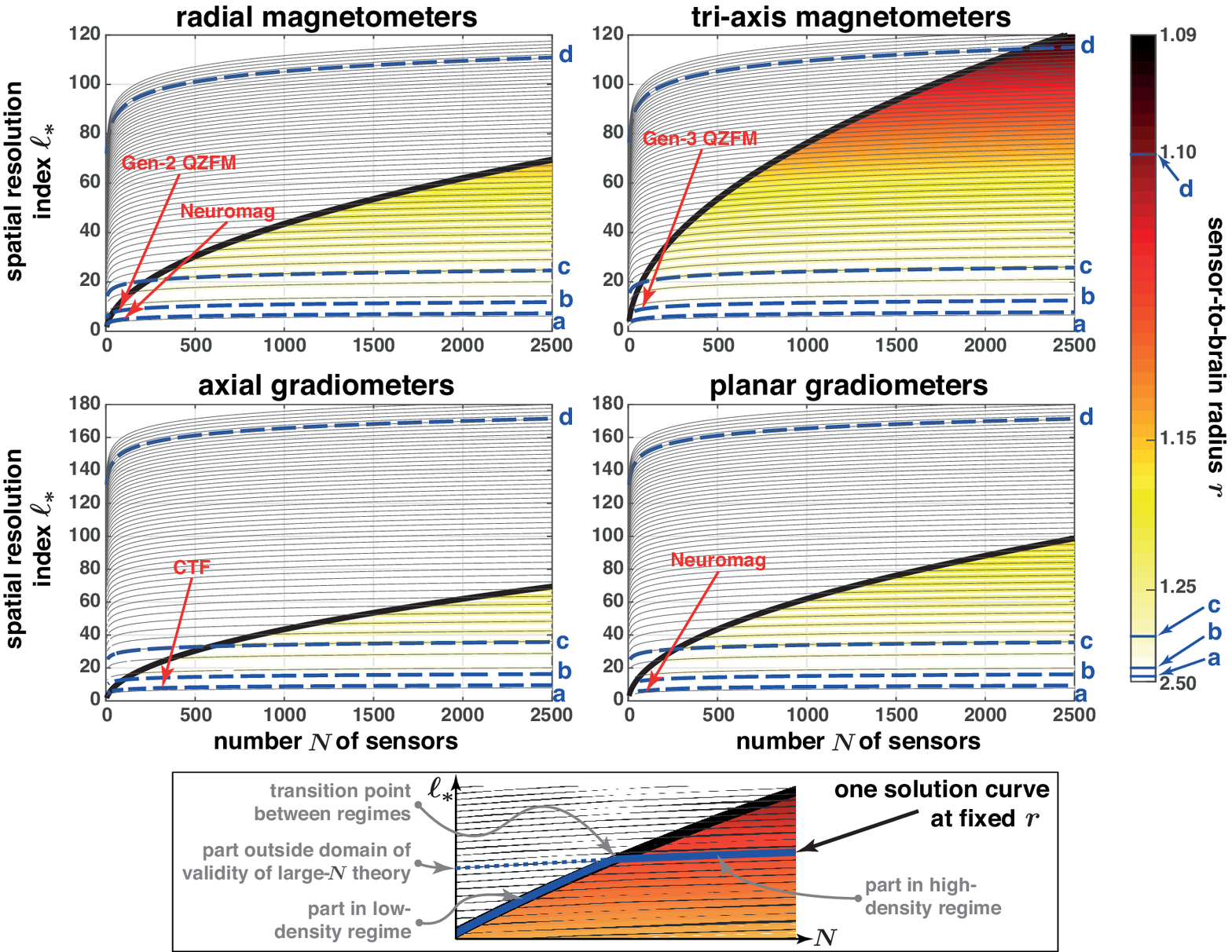}
\caption{\label{fig4} \emph{Limits of spatial resolution for hemispherical MEG arrays.} The spatial resolution index $\ell_*$ is plotted as a function of the number $N$ of sensors based on the asymptotic theory (Eq.~\ref{lstar-subleading}), for different sensor types (\textbf{top left}: radial magnetometers, \textbf{top right}: tri-axis magnetometers, \textbf{bottom left}: axial gradiometers, \textbf{bottom right}: planar gradiometers), various values of sensor-to-brain distances ($1.09\leq r\leq 2.50$; each value corresponding to a light grey curve), and the multipole SNR set to $\sigma_{\boldsymbol a}/\sigma_{\boldsymbol \varepsilon}=12.5$ (Table \ref{Table4}). The domain of validity of the asymptotic theory is color-coded according to the sensor-to-brain distance $r$, with the four estimates reported in Table \ref{Table4} further emphasized (blue dashed curves; \textbf{a}: functional cryogenic MEG, \textbf{b}: functional scalp MEG, \textbf{c}: anatomical cryogenic MEG, \textbf{d}: anatomical scalp MEG). The part of the large-$N$ solutions outside this domain must be replaced by the solution in the linear regime (thick black curve; see Eq.~\ref{lstar-linear}) as illustrated in the bottom insert.
The working points of existing MEG systems are indicated by arrows (Neuromag: $N=102$; CTF: $N=275$; scalp MEG with QZFM: $N\approx 50$). QZFM: quantum zero-field magnetometer (QuSpin; Gen-2 corresponds to single-axis radial magnetometers and Gen-3, to tri-axis magnetometers).
}
\end{figure*}

Based on the previous data, we provide in Fig.~\ref{fig4} a comprehensive view of the spatial resolution index $\ell_*$ for hemispherical MEG arrays with any number $N$ of sensors (here limited to $N\leq 2500$) and for a range of sensor-to-brain distances encompassing the four cases analyzed above. The domain of validity of the asymptotic theory described by Eq.~\eqref{lstar-subleading} is emphasized by the colored area in Fig.~\ref{fig4}. Its boundary (black thick line) corresponds to the transition into the low-density regime and was described by the parametric curve
\begin{equation}\label{lstar-linear} \ell_* = -3/2 + \sqrt{MN+1/4}\, \qquad \textrm{(low-density regime)}\, , \end{equation}
which is nothing but the linear description \eqref{nu-linear} of signal-space dimension translated into the spatial resolution index via Eq.~\eqref{nu-hemisphere}. We extrapolated the spatial resolution index to the low-density regime using Eq.~\eqref{lstar-linear}, notwithstanding the overestimation that this gluing procedure entails around the transition (Fig.~\ref{fig3}).
This means that all large-$N$ solution curves in Fig.~\ref{fig4} (colored area) reaching the transition from the right (i.e., decreasing $N$) collapse and follow the transition line (black thick line) rather than pass through to its left (Fig.~\ref{fig4}, bottom insert). We check the consistency of the low-density ansatz \eqref{lstar-linear} with a physical description of the transition in \ref{appendix-smoothness}.

%

Figure \ref{fig4} confirms that adding sensors increases the spatial resolution index, rapidly while in the low-density regime (black thick curve) and then much more slowly once the high-density regime is reached (colored area). It is interesting that the working points of functional MEG (inferred from resting-state recordings) in existing cryogenic systems were similar (Neuromag, $\ell_*=7.2$ for $N=102$ planar gradiometers; CTF, $\ell_*=7.8$ for $N=275$ axial gradiometers; Table \ref{Table2}) and both within the high-density regime (Fig.~\ref{fig4}; see arrows pointing to curves a), so adding sensors would hardly improve their spatial resolution. This was illustrated by the flatness of the respective solution curves beyond their working points (Fig.~\ref{fig4}, bottom, curves a). In fact, augmenting sensor density to unrealistic levels led to modest gains (Neuromag, $\ell_*=9.3$ for $N=2500$ planar gradiometers; CTF, $\ell_*=9.5$ for $N=2500$ axial gradiometers).
More surprising is the observation that functional MEG on scalp already stood in the high-density regime too, even at the relatively low number ($N\approx 50$) of sensors available in current OPM systems (Fig.~\ref{fig4}, top; see arrows pointing to curves b). In fact, the spatial resolution of OPM recordings at rest with 50 tri-axis magnetometers ($\ell_*=8.5$; see Table \ref{Table2}) already outperformed resting-state cryogenic MEG ($\ell_*\leq 7.8$). Realistic augmentations of OPM density would lead to limited improvements ($\ell_*=9.2$ for $N=102$ tri-axis magnetometers; see Table \ref{Table2}), and even reaching unrealistic densities would not revolutionize the situation ($\ell_*=12.5$ for $N=2500$; see Fig.~\ref{fig4}, top right, curve b). This is all reminiscent of the logarithmic slowness described by \vince{Eq.~\eqref{lstar-subleading} (see also Eq.~\ref{lstar-leading} and \ref{appendix-proofs-properties})}. We may thus conclude that \emph{further developing MEG sensor technology towards denser arrays is not an efficient way to improve MEG spatial resolution}.

\begin{table*}[ht]
\begin{center}
\begin{tabu}{|[1pt]c|[0.5pt]c|[1pt]c|[0.5pt]c|[0.5pt]c|[0.5pt]c|[1pt]}
    \tabucline[1pt]{1-6}
    \textbf{system} & \textbf{sensor type} & \textbf{func-cryo\textsuperscript{(a)}} & \textbf{func-scalp\textsuperscript{(b)}} & \textbf{anat-cryo\textsuperscript{(c)}} & \textbf{anat-scalp\textsuperscript{(d)}} \\
    \tabucline[1pt]{1-6}
    \multirow{2}{*}{Neuromag} & $N=102$ radial magn & $\boldsymbol{\ell_*=4.9}$ & $\ell_*=8.6$ & $\ell_*=12.8^\dagger$ & $\ell_*=12.8^\dagger$ \\ [-1pt] \tabucline[0.5pt]{2-6} & $N=102$ planar grad & $\boldsymbol{\ell_*=7.2}$ & $\ell_*=12.6$ & $\ell_*=18.8^\dagger$ & $\ell_*=18.8^\dagger$ \\
    \tabucline[1pt]{1-6}
    \multirow{1}{*}{CTF} & $N=275$ axial grad & $\boldsymbol{\ell_*=7.8}$ & $\ell_*=13.9$ & $\ell_*=22.0^\dagger$ & $\ell_*=22.0^\dagger$ \\
    \tabucline[1pt]{1-6}
    \multirow{2}{*}{Gen-3 QZFM} & $N=50$ tri-axis magn & $\ell_*=4.8$ & $\boldsymbol{\ell_*=8.5}$ & $\ell_*=15.8^\dagger$ & $\ell_*=15.8^\dagger$ \\ [-1pt] \tabucline[0.5pt]{2-6} & $N=102$ tri-axis magn & $\ell_*=5.4$ & $\ell_*=9.2$ & $\ell_*=20.2$ & $\ell_*=23.3^\dagger$ \\
    \tabucline[1pt]{1-6}
    \end{tabu}
\end{center}
\caption{\label{Table2}{\it Spatial resolution index $\ell_*$ for selected MEG systems.} Values were extracted from solution curves at four sensor-to-brain distances in the high-density regime (Fig.~\ref{fig4}, curves \textbf{a}--\textbf{d} in colored area) or in the low-density regime (Fig.~\ref{fig4}, thick black line), for Neuromag-like MEG (102 radial magnetometers and 102 planar gradiometers), CTF-like MEG (275 axial gradiometers), and Gen-3 QZFM MEG (50 or 102 tri-axis magnetometers). The cases corresponding to current experimental MEG systems are emphasized in bold. func/anat: functional/anatomical parameter estimates (Table \ref{Table4}), cryo/scalp: cryogenic/scalp MEG, magn: magnetometer, grad: gradiometer, QZFM: quantum zero-field magnetometer, $\dagger$: prediction taken from the low-density regime (Eq.~\ref{lstar-linear}).}
\end{table*}

The second way to improve MEG spatial resolution is by reducing the gap between sensors and sources of brain activity. This translates in Fig.~\ref{fig4} by the increasing elevation of solution curves as the sensor-to-brain distance decreased. For example, moving a cryogenic MEG on scalp in resting-state recordings (functional MEG) corresponds to passing from curves (a) to curves (b) in Fig.~\ref{fig4} and led to a gain of $70$--$78\%$ in the spatial resolution index (see Table \ref{Table2}, columns 3 and 4). This improvement is moderate because the functional brain sphere inferred from resting-state MEG was smaller than what anatomy entails (Table \ref{Table4}), effectively leaving a substantial gap between scalp and brain sources. Our asymptotic theory predicted that this effect would become enormous as the sensors are brought to the vicinity of brain sources (since the spatial resolution index diverges as the sensor-to-brain distance $r$ approaches 1, see Eq.~\ref{lstar-subleading} and \ref{appendix-proofs-properties}), as shown very clearly by the solution curves for anatomical scalp MEG (Fig.~\ref{fig4}, curves d). However, the effect was strongly mitigated in practice because the elevation of solution curves must saturate to the height of the transition line (Fig.~\ref{fig4}, black thick curve). For example, all curves (d) in Fig.~\ref{fig4} stood well within the low-density regime ($N\leq 2223$), so the spatial resolution index actually saturated at the values dictated by the low-density regime (Eq.~\ref{lstar-linear}). We conclude that \emph{spatial resolution saturates when MEG sensors are close enough to brain sources to a maximum value controlled by the total number $MN$ of recording channels}. The gain in spatial resolution index for cryogenic MEG may not exceed $160\%$ for Neuromag resting-state recordings ($\ell_*=18.8$; Table \ref{Table2}, columns 3 and 6), $180\%$ for CTF ($\ell_*=22.0$), and $230\%$ for an OPM array with 50 tri-axis magnetometers ($\ell_*=15.8$; Table \ref{Table2}, columns 4 and 6).

Third and last, Fig.~\ref{fig4} confirms that gradiometers lead to higher spatial resolution than magnetometers (compare the scales in the top and bottom panels). Of note, solution curves in the high-density regime (Fig.~\ref{fig4}, colored areas) were nearly indisinguishable between radial and tri-axis magnetometers, and between axial and planar gradiometers. That is in line with the lack of impact of sensor components disclosed by our asymptotic theory. 
However, the number $M$ of sensor components plays an important role in determining the transition between low- and high-density (Eq.~\ref{lstar-linear} and Fig.~\ref{fig4}, thick black curve). \emph{The main advantage of multi-component MEG sensors (for spatial resolution) is to extend the domain of validity of the asymptotically high-density regime.} This means that the spatial resolution index is higher for tri-axis magnetometers than for single-axis magnetometers at reasonably small number $N$ of sensors (see also Table \ref{Table2}).

\newpage

\section{Discussion}\label{Discussion}

We presented a detailed analysis of MEG spatial resolution based on multipolar expansions, with a particular emphasis on the limits of MEG spatial resolution. We developed an asymptotic theory describing these limits analytically, and used experimental and numerical data to investigate under what conditions the theory breaks down. This combined approach led to a characterization of MEG spatial resolution in terms of two qualitatively distinct regimes. First, the high-density regime corresponds to the domain where the number of sensors is large enough for MEG spatial resolution to be faithfully described by the asymptotic theory. It featured (i) very slow, yet unbounded, improvements when adding sensors to MEG arrays, (ii) rapid, but bounded, improvements as MEG sensors are brought closer to the scalp, and (iii) moderately higher spatial resolution for gradiometers than magnetometers (independently of sensor components and SNR). Second, the low-density regime corresponds to the breakdown of the asymptotic theory when the number of sensors is too small. It featured much faster spatial resolution gains when adding sensors, and these gains were enhanced with multi-component sensors (e.g., planar gradiometers or tri-axis magnetometers). The interplay between these two regimes controls the saturation value of MEG spatial resolution as sensors approach brain sources.

\subsection{The physics of MEG spatial resolution}\label{discuss-megphys}

\paragraph{Magnetic field smoothness controls spatial resolution} We investigated a specific notion of MEG spatial resolution, i.e., the focality and smoothness of neuromagnetic topographies measured by the largest angular frequency accessible to noisy multipolar expansions. Our inspiration came from the works surrounding signal-space separation \citep{Taulu2004,Taulu2005} but this approach is close in spirit to previous theoretical studies of MEG spatial resolution \citep{Ahonen1993,Iivanainen2021}, and most particularly to \cite{Tierney2022} who explored OPM spatial sampling properties with simulations of MEG multipolar expansion models. Other metrics of spatial resolution comprise leadfield focality \citep{Boto2016} as well as source localization accuracy/separability \citep{LucenaGomez2021,Sekihara2005,Vrba2004} and cross-talk/point-spread functions \citep{Hauk2014,Wens2015}, although the latter techniques actually mix MEG spatial resolution \emph{per se} with the spatial smoothness imposed by the choice of inverse model. Other metrics often used in the literature are signal complexity measures such as signal-space dimension (that we used here; see Eq.~\ref{sdof-a}; see also Section \ref{Methods} for a version based on MEG forward modeling), leadfield rank \citep{Tierney2020}, or total information \citep[][]{Ilvanainen2017}. However, our theoretical results highlighted the fact that these metrics mix spatial resolution with geometric aspects such as the amount of coverage. Notwithstanding, our results largely agree with previous studies \citep{Ahonen1993,Boto2016,Ilvanainen2017,Iivanainen2021,Marhl2022,Tierney2020,Tierney2022,Vrba2004}; what we bring is a new perspective explicitly focused on the physics of neuromagnetic fields and the limit of asymptotically high sensor density.


The physical process that inherently limits MEG spatial resolution is magnetic field smoothness. In terms of multipolar expansions at the basis of our analysis, it corresponds to the exponential suppression of focal neuromagnetic topographies at high angular frequencies. That explicit representation of field smoothness is precisely what allowed us to measure MEG spatial resolution in terms of angular frequency. In this sense, our spatial resolution index is directly controlled by the physics of magnetic field smoothness.

\paragraph{Spatial resolution is limitless} One of our main conceptual result is the new finding that adding sensors to a MEG array improves spatial resolution endlessly, albeit these improvements bear an increasingly high cost. That was contrary to our initial hypothesis that MEG spatial resolution would converge to a definite limit. Intuitively, packing further sensors in a MEG array where sensor separation is already well below the typical ``blur size'' associated with field smoothness should not add new information, analogously to EEG electrode bridging by electrolyte spread.
In a sense, we assumed that field smoothing would act as a hard low-pass spatial filter on MEG topographies. In hindsight, this intuition was wrong because this smoothing implements a physical, analog spatial filter and must therefore be soft, i.e., it suppresses the contribution of small spatial scales but does not eliminate them altogether. That is why it is, in principle, possible to probe more and more focal patterns with sufficiently refined sensor grids. Still, this theoretical finding bears no dramatic consequences for MEG practice because the increase in spatial resolution is slow; not only slow, logarithmically slow! In that sense, magnetic field smoothness does constraint MEG spatial resolution.

On a side note, the theoretical ability of MEG to harvest infinite amounts of information does not mean that the infamous inverse problem \citep{Hamalainen1993MEG} may be solved by increasing sensor density. A hypothetical MEG array composed of a continuum of sensors would still be blind to electrical source configurations that are magnetically silent \citep{Hamalainen1993MEG}. The inverse problem is an issue related to the non-invasive nature of MEG, not to its spatial resolution \emph{per se}.

\paragraph{Spatial resolution exhibits two qualitative regimes} The asymptotic behaviour of MEG spatial resolution, particularly its slow divergence, determined what we called the high-density regime. Physically, this corresponds to situations where the separation between neighboring sensors is well below the size of magnetic field smoothness. (See \ref{appendix-smoothness} for an order-of-magnitude characterization of smoothness size.)
It is unsurprising that an entirely different behaviour emerged is the opposite, low-density regime wherein sensor separation is larger than field smoothness. In this situation, the intrinsic smoothness of neuromagnetic patterns does not impact sensors, which bring independent informations. Spatial resolution is thus controlled by the sheer number of recording channels and increases as fast as possible as sensors are added to a low-density MEG array. 
This is in line with previous work \citep[e.g.,][]{Tierney2022,Vrba2004}.
Further, the number of components recorded per sensor plays an important role at low density by controlling spatial resolution gains. This illustrates one crucial advantage of using, e.g., tri-axis magnetometers instead of single-axis radial magnetometers for OPM system designs with limited amounts of sensors \citep[notwithstanding their added-value for OPM denoising;][]{Brookes2021}.

The usefulness of multi-component sensors was notably absent in the high-density regime. Physically, measurements of different magnetic components ($\boldsymbol B$ or its gradient) at infinitesimally close sensor locations are necessarily inter-related (via the equations of magnetostatics, $\boldsymbol \nabla \times \boldsymbol B=0$ and $\boldsymbol \nabla \cdot \boldsymbol B=0$). In fact, we observed that the limits of MEG spatial resolution are the same for different types of magnetometers (radial vs.~tri-axis) or gradiometers (axial vs.~planar). Such convergence was already reported by \cite{Tierney2022} for scalp magnetometers and by \cite{Ahonen1993} for gradiometers; our results gather these observation within a single theoretical framework. On the other hand, the magnetometric or gradiometric nature of the MEG array does impact the limits of MEG spatial resolution, with gradiometers showing higher resolution. This simply reflects the fact that the sensitivity profile of gradiometers is equivalent to that of two neighboring magnetometers with opposite orientation. This double magnetometer configuration turns out to improve spatial resolution by more than a mere doubling of the number of magnetometers, because sensor separation is not homogeneous but rather is tuned to sense fine topographical details. In an interesting twist, the low-density regime was impervious to the magnetometric or gradiometric nature of sensors.

Perhaps more surprisingly, combining these two distinct, \vince{analytically tractable} regimes allowed for a reasonable description of MEG spatial resolution as a whole. Our data showed that spatial resolution was overestimated around the transition between low- to high-density regimes. Why our two-regime picture of spatial resolution led to an overestimation is explained by the fact that the low-density description neglects completely the effects of magnetic field spread and of cross-talk across components, both of which introduce inter-dependencies amongst sensors and thus reduce spatial resolution. Modeling this in-between situation analytically represents an interesting challenge to refine our understanding of MEG spatial resolution at the transition.

\paragraph{Reducing array size improves spatial resolution by effectively lowering sensor density} Our analysis confirmed the expected result that bringing sensors on scalp improves MEG spatial resolution, i.e., the extent of magnetic field smoothness decreases. The higher focality of neuromagnetic topographies obtained with scalp MEG (compared to cryogenic MEG) is well known and actually provided an essential initial motivation for the development of OPM-based MEG \citep{Boto2016}. Still, our two-regime description of spatial resolution sheds new light on the detailed physics underlying this seemingly trivial aspect.

We showed that approaching sensors to the scalp increases spatial resolution by moving the system from the high- to the low-density regime, until spatial resolution saturates to a maximum value once the low-density regime is reached. This saturation effect has not been described before and was, in our opinion, puzzling and counter-intuitive. Shrinking a MEG array increases sensor density so it should bring it away from the low-density regime, not into it! The solution to this puzzle lies in the interplay between geometry and the physics of magnetic field smoothness; what matters for MEG spatial resolution is sensor separation relative to the size of field smoothness. The geometric separation between neighboring sensors obviously decreases as the MEG array shrinks. The extent of field smoothness decreases concomittantly as sensors get closer to the sources of brain activity, but faster than geometric distances (\ref{appendix-smoothness}). This results into an augmentation of sensor separation, and thus a diminution of sensor density, when they are expressed in units of field smoothness size. That is precisely what moving the system towards the low-density regime means. (We refer to \ref{appendix-smoothness} for a formal version of the above argument.) Once the low-density regime is reached, the relative sensor separation is large enough that sensors are insensitive to field smoothness and thus are independent. Further shrinking will continue to augment the relative separation but this cannot render sensors more independent, which explains why spatial resolution saturates.

\paragraph{Spatial resolution depends on brain activity} The effect of sensor-to-brain distance warrants further discussion on what we meant by ``brain size'' in our analysis. Using the literal size of the anatomical brain led to maximal, fully saturated spatial resolution for scalp MEG (i.e., in the low-density regime whatever the number of sensors) because the brain surface is very close to the scalp. Generally speaking, this saturation is likely an artifact rather than a reflection of reality, as we showed using experimental MEG data. The maximum-entropy hypothesis (that was needed to develop the theory) entails the unphysical assumption that all possible source configurations are equally active within the brain sphere. This includes activations in neocortical areas of the brain convexity just beneath the scalp. Such activations would dominate scalp MEG recordings and lead to highly focal topographies at the nearest sensors. In turn, the spatial smoothness associated to this dominating activity would be so small that MEG spatial resolution saturates virtually whatever the (geometric) sensor density. Clearly, this does not correspond to functional brain physiology. Although similarly unrealistic assumptions are used very successfully in MEG source projection techniques \citep[such as minimum-norm estimation; see][see also Section \ref{param-func}]{Dale1993}, the consequences for MEG multipolar expansions appeared more problematic.

Our way to deal with this issue was to replace brain anatomy with a functional estimate of brain size, using resting-state MEG activity as prototypical example. A functional brain sphere was determined as the equivalent brain sphere that reproduces the information content (and thus the spatial resolution) of MEG recordings while indulging the unrealistic maximum-entropy hypothesis, i.e., that all source configurations within this equivalent sphere are equally active. Its size thus corresponds in a sense to an average weighted by the amount of neural activity in neocortical and deep cortical regions. The functional brain sphere inferred from cryogenic MEG resting state was thus, of necessity, smaller than the anatomical brain. This led to realistic levels of spatial resolution and notably moderate improvements when passing from cryogenic to scalp MEG. In general though, the type of brain activity under study may affect the functional brain size and thus MEG spatial resolution. For example, focusing on high-SNR events, such as interictal epileptiform spike-wave discharges occurring near the neocortical surface under the skull, is bound to enlarge the functional sphere, further close the gap with sensors, and increase spatial resolution. The anatomical MEG model considered in our numerical analyses actually reflect such cases of optimal spatial resolution. This illustrates the enormous potential of scalp MEG for the clinical diagnosis of epilepsy \citep{Feys2022, Vivekananda2020, Widjaja2022}. Contrariwise, high-SNR activity in deep brain regions, such as hippocampal epileptic discharges, would correspond to a small functional brain sphere and thus to both poor spatial resolution and limited improvements brought by scalp MEG \citep[notwithstanding imaginative uses of OPMs to probe deep brain activity; see][]{Tierney2021}.


The functional dependence of MEG spatial resolution, along with the saturation effect, has another interesting implication for clinical MEG in epilepsy. The gold standard for high-resolution electrophysiological mapping of interictal epileptiform spike-wave discharges currently remains invasive techniques such as electrocorticography \citep[ECoG;][]{Jost2020}. In this context, one might wonder whether a hypothetical magnetic version of ECoG, i.e., magnetocorticography, would lead to further improvements. Simulations suggest that scalp MEG might eventually outperform ECoG \citep{Nugent2022}, opening the possibility that the former replace the latter in the future. The remaining question is whether placing sensors directly on the neocortical surface (notwithstanding technical feasibility) would lead to even higher-resolution recordings of epilepsy. The above discussion suggests a negative answer, at least for neocortical epilepsies, because anatomical MEG spatial resolution (which models the situation of epileptiform spike-wave discharges right under the neocortical convexity) already saturates on scalp to its maximal value and thus cannot improve by invasive recordings. Although this line of thoughts would benefit from experimental validation, it adds to the long list of arguments suggesting that scalp MEG based on OPMs may lead, in time, to a step change in the clinical diagnosis of neocortical epilepsy \citep{Feys2022, Widjaja2022}.

The dependence of MEG spatial resolution in brain activity further raises a couple of noteworthy observations. First, this dependence arises through the way neural sources are distributed across the cortex, not directly through their SNR. Our analysis highlighted that spatial resolution is actually quite impervious to the multipole SNR parameter, i.e., a global measure of source SNR. This claim may appear surprising at first given how the point-spread function of adaptive MEG source projection algorithms such as beamforming vary with the SNR level \citep{VanVeen1997}; what this merely indicates is that beamformer point-spread functions do not provide a faithful measurement of spatial resolution. Second, this functional dependence implies that MEG spatial resolution must actually be a dynamic parameter that evolves alongside source activity. Our current approach to functional parameter estimation overlooked this aspect, so it would be interesting in future developments to try and include temporal information.

\subsection{Implications for MEG technology}\label{discuss-practice}

\paragraph{The race for extremely high-density MEG is costly}
Even though spatial resolution may in theory always be improved by adding sensors, in practice it is increasingly difficult to obtain substantial improvements. A trade-off must therefore be found based on different considerations about MEG system design, among which the cost of spatial resolution. We can actually quantify this cost (\ref{appendix-proofs-properties}); it scales with the number $N$ of sensors as $\sqrt{N/M}$ in the low-density regime and as $\log r\times N$ in the high-density regime (with $M$ the number of sensor components and $r$ the sensor-to-brain distance). These different scalings embody quantitatively the fact that improving spatial resolution by adding sensors is much less costly at low density than at high density. Recording multiple field components further decreases cost in the low-density regime but not in the high-density regime. Approaching sensors to the scalp reduces cost in the high-density regime but not in the low-density regime. Combining these considerations with other aspects of production (e.g., cost per sensor, sensor size, or array weight) might be helpful for the decision-making process taking place during the design of novel MEG systems. That process is nowadays largely moot for cryogenic MEG (for which we showed that their working point implements a trade-off), but it is vividly ongoing for scalp MEG based on the developing OPM technology \citep{Hill2020,Ilvanainen2017,Iivanainen2021,Tierney2020,Tierney2022}. In this regard, it is interesting that the working point of current state-of-the-art OPM systems (about 50 tri-axis magnetometers) already stood in the high-density regime despite the limited amount of sensors compared to cryogenic MEG. This means that pursuing the race of increasingly denser OPM arrays may not be the most efficient way to improve the performance of scalp MEG. This conclusion is in agreement with the analysis of OPM spatial sampling by \cite{Tierney2020,Tierney2022}.

This being said, spatial resolution represents only one aspect of MEG performance. In particular, the benefits of spatial oversampling cannot be overstated as it allows efficient signal denoising. Oversampling corresponds to the situation where there are many more recording channels than degrees of freedom in a MEG array. Cryogenic MEG systems (Neuromag and CTF) are oversampled as their working point lies deep enough within the high-density regime (\ref{appendix-proofs-properties}), where existing interference suppression algorithms relying on oversampling can be extremely efficient \citep{Larson2018,Taulu2004,Taulu2005}. The situation of scalp MEG is not as confortable yet \citep{Seymour2022}; current OPM arrays are not oversampled because their working point lies too close to the low-density regime (\ref{appendix-proofs-properties}). Reaching oversampling will require further miniaturization of OPM sensors and the suppression of cross-talk among neighboring sensors generated by onboard field zeroing coils \citep{Nardelli2019b}. In the meantime, the emergence of tri-axis OPMs may help and ameliorate the suppression of both external magnetic interferences and movement artefacts in scalp MEG \citep{Brookes2021}.

\paragraph{The spatial resolution index may be used to benchmark MEG designs}
The above discussion suggests that our data, particularly Fig.~\ref{fig4}, might find useful applications for benchmarking the design of MEG arrays. This is particularly relevant in the current context of frenetic OPM developments, which rely on both simulation studies \citep{Iivanainen2021,Tierney2020} and statistical frameworks such as the minimization of source reconstruction error bounds \citep{Beltrachini2021,Muravchik2001} or Bayesian model comparison of sensor array geometries \citep{Duque-Munoz2019}. In fact, it would be interesting in future works to compare data-driven approaches to OPM array optimization with our analysis that specifically assesses their spatial resolution. That being said, our analytical theory exhibits a few limitations that should be kept in mind in such endeavours.

First and foremost, our analysis focused on fairly idealized measurements of the neuromagnetic field. One hidden but major assumption behind the observation model \eqref{obsmodel} is that experimental measures coincide with actual point field values in the absence of noise. This means that sensors exhibit no cross talk, perfect calibration and geometry, linear gains, and their spatial extent (i.e., a SQUID measures the magnetic flux through its pick-up coil and an OPM, a field component averaged within its vapor cell) is negligible. These imperfections are generally minimized at the MEG hardware level \citep[see, e.g.,][in the context of OPMs]{Holmes2018,Holmes2019,Tierney2019}, but taking them into account may be critical to reach optimal accuracy \citep[see, e.g., Maxwell filtering for SQUID-based MEG;][]{Taulu2004, Taulu2005}. On the other hand, we did include sensor noise (modeled as homogeneous and uncorrelated across sensors) explicitly in our description. That was, in fact, instrumental in our determination of MEG spatial resolution as it depends on the balance between magnetic field spread and noise level (although, surprisingly, the precise noise level does not impact spatial resolution estimation). However, we neglected completely external interferences from the background magnetic environment and from head movements (either in a fixed helmet for cryogenic MEG or with sensors moving in the background field for scalp MEG). 
This being said, MEG design optimisation along with the usage of interference suppression algorithms should mitigate this possible overestimation.

Second, we considered an idealized MEG array geometry with hemispherical sensor coverage. That is not fundamentally a bad approximation \citep[see, e.g., the scalp OPM cap presented in][]{Feys2022}, although whole-head MEG tends to cover a larger scalp area to better sample temporal regions (see, e.g., Fig.~\ref{fig1}). Interestingly, the spatial resolution index turned out identical for hemispherical MEG and spherical MEG, i.e., an idealized array fully surrounding the brain (\ref{appendix-proofs-spectrum}), so we surmise that sensor coverage does not affect our spatial resolution measure. We found that it does change signal-space dimension $\nu$ though; it was approximately twice larger for fully spherical MEG. On this basis, we conjecture that it scales with the coverage area $a$ relative to the hemisphere (between $a=1$ for an hemisphere $\mathbb H$ and $a=2$ for the whole sphere $\mathbb S$; e.g., $a=1.15$ for the Neuromag array), i.e., $\nu\approx a\, \nu_\mathbb{H}$. It is noteworthy that this simple rule slightly underestimates the actual signal-space dimension (\ref{appendix-proofs-properties}); physically, magnetic field smoothness effectively extends the field of view beyond the borders of the array. \vince{Also noteworthy is} that our data comparing Neuromag-like MEG arrays and hemispherical MEG arrays were fully compatible with this rule of proportionality. This supports both this rule and the idea that MEG spatial resolution does not depend on the precise shape of sensor arrays. What matters though is the assumed homogeneity of sensor coverage, as indicated by the observation that replacing homogeneously distributed magnetometers by gradiometers (which corresponds to a highly inhomogeneous distribution of twice as many magnetometers) boosts spatial resolution substantially more than a mere doubling of magnetometer density.

Despite the idealized nature of our description of MEG spatial resolution, we expect it to provide realistic estimates. As a proof of concept, let us emphasize our finding that the spatial resolution index was $\ell_*=7.2$ for Neuromag MEG recordings at rest. This is consistent with our preprocessing based on signal-space separation, which truncated interior MEG multipolar expansions to angular frequencies below $\ell_\textrm{in}=8$ so that our spatial resolution index was constrained from the start to $\ell_*\leq \ell_\textrm{in}$. Still, our asymptotic theory together with the resting-state functional estimate of brain size was able to recover this parameter \citep[which was determined by different means in][]{Taulu2004,Taulu2005} instead of finding substantially lower values. We envision that our results may be useful not only to benchmark hardware MEG designs but also to fine tune denoising parameters in future OPM applications of signal-space separation \citep{Seymour2022}. It should be further possible to extend our analysis and include exterior multipolar expansions to model background magnetic fields and estimate the external truncation parameter $\ell_\textrm{out}$ as well \citep{Taulu2004,Taulu2005,Tierney2022}.

As another side-product, our theory may find useful applications in the mass-univariate statistical analysis of MEG spatial maps. Existing techniques rely on heuristic approaches to estimate the number $\nu$ of spatial degrees of freedom available in MEG maps, and use this number as Bonferroni correction factor to control the false positive rate \citep{Barnes2011,Wens2015}. Our results on signal-space dimension $\nu\approx a\, \nu_{\mathbb H}$ (i.e., including the sensor coverage factor $a$ discussed above) provides an analytically grounded and rigorous basis for such approaches, so it might play a role somewhat analogous to random field theory in the statistical analysis of functional MRI or positron emission tomography \citep{Worsley1996}. 

In conclusion, our two-regime theoretical model of MEG spatial resolution allows not only for better insights into the physics of neuromagnetism, but also for diverse applications that may help both hardware and software optimisations of scalp MEG based on the rising OPM technology.

%
%
%
%
%
%
%


\section*{Acknowledgements}

I would like to thank Dr.~Nicolas Coquelet for sharing MEG data and Pr.~Xavier De Ti\`ege for valuable comments on this manuscript. \vince{This work was supported by the F.R.S. -- FNRS (research convention Excellence of Science EOS MEMODYN, 30446199).} The MEG project at the H.U.B. -- H\^opital Erasme is financially supported by the Fonds Erasme (research conventions ``Les Voies du Savoir'' for cryogenic MEG and ``Projet de recherche clinique \`a des techniques m\'edicales \'emergentes 2020'' for scalp MEG based on OPMs).


\appendix
\setcounter{table}{0}
\setcounter{figure}{0}

\section{The machinery of MEG multipolar expansions: Review and minor results}\label{appendix-bckgd}
\setcounter{table}{0}
\setcounter{figure}{0}

We review here useful background on multipolar expansions and spherical harmonics on the sphere. We also include minor new results associated to the maximum-entropy hypothesis \eqref{maxentropy} and to hemispherical multipolar expansions, which are used in the main text and in \ref{appendix-proofs}.

\subsection{A brief review of MEG multipolar expansions}\label{appendix-bckgd-MPEreview}

\paragraph{Magnetostatic potential} The building block of MEG multipolar expansion models is the notion of magnetostatic potential $U$ \citep{Jackson,Zangwill2012}. Since neuromagnetic activity is probed outside the head and works in a quasi-static regime, the magnetic field $\boldsymbol B$ in a neighbourhood of the array can be expressed as the gradient \citep{Hamalainen1993MEG}
\begin{equation}\label{potential} \boldsymbol B = \boldsymbol \nabla U\, . \end{equation}
This actually holds at any location outside the smallest sphere enclosing the brain (shown blue in Fig.~1), i.e., as long as the radial distance to the sphere center exceeds the brain sphere radius $R_\textrm{brain}$. The potential $U$ in this extra-cranial domain satisfies Laplace's equation; it can thus be subjected to an interior multipolar expansion
\begin{equation}\label{MPE-potential} U(r,\Omega) = \sum_{\ell,m} a_{\ell,m}\, \frac{Y_{\ell,m}(\Omega)}{r^{\ell+1}}\end{equation}
over the spherical harmonics $Y_{\ell,m}$ indexed by integers $\ell\geq 0$ and $-\ell\leq m\leq \ell$ \citep[][see also \ref{appendix-bckgd-MPEreviewYlm}]{Jackson,Zangwill2012}. The expansion parameter is taken here as the inverse of the relative radial coordinate $r=R/R_\textrm{brain}$, so $r>1$ in Eq.~\eqref{MPE-potential}.

\paragraph{Field components and derivatives}
Different field measurements $\boldsymbol \phi$ can then be examined by taking the gradient \eqref{potential} of Eq.~\eqref{MPE-potential} and extracting the relevant components or derivatives. The cases of interest for current MEG technology are listed in Table \ref{TableA1}.
\begin{table}[ht]
\begin{center}
\setlength{\tabulinesep}{2pt}
\begin{tabu}{|[1pt]c|[1.5pt]c|[1pt]c|[1pt]c|[1pt]c|[1pt]}
    \tabucline[1pt]{1-4}
    \textbf{sensor type} & $M$ & $\boldsymbol \phi$ & $(\phi_1, \ldots, \phi_M)$  \\
    \tabucline[1.5pt]{1-4}
    radial magn & 1 & $\boldsymbol n \cdot \boldsymbol B$ & $B_r$ \\
    \tabucline[1pt]{1-4}
    planar magn & 2 & $\boldsymbol n \times \boldsymbol B$ & $(B_\theta,B_\varphi)$ \\
    \tabucline[1pt]{1-4}
     tri-axis magn & 3 & $\boldsymbol B$ & $ (B_r,B_\theta,B_\varphi)$ \\
    \tabucline[1pt]{1-4}
     axial grad & 1 & $\boldsymbol n \cdot \boldsymbol \nabla (\boldsymbol n \cdot \boldsymbol B)$ & $\frac{\partial B_r}{\partial r}$ \\
    \tabucline[1pt]{1-4}
     planar grad & 2 & $\boldsymbol n \times \boldsymbol \nabla (\boldsymbol n \cdot \boldsymbol B)$ & $\left( \frac{1}{r} \frac{\partial B_r}{\partial \theta},  \frac{1}{r \sin\theta} \frac{\partial B_r}{\partial \varphi} \right)$ \\
    \tabucline[1pt]{1-4}
    \end{tabu}
\end{center}
\caption{\label{TableA1}{{\it Definition of the $M$-vector $\boldsymbol \phi$ for different sensor types.}} The unit vector $\boldsymbol n$ denotes the outward normal to the array surface at location $\Omega$. The dot product with $\boldsymbol n$ extracts the normal component of the magnetic field $\boldsymbol B$ or the gradient operator $\boldsymbol \nabla$, whereas the cross product with $\boldsymbol n$ projects on the tangent plane. Note that the two tangential components always appear in combination. Components $\phi_i$ are also written with spherical coordinates and $\boldsymbol n$ pointing radially outwards (Fig.~1). magn: magnetometer, grad: gradiometer.}
\end{table}
These expressions can be used to compute the components $S_i(\Omega | \ell,m)$ ($1 \leq i \leq M$) of the vectorial spherical harmonics $\boldsymbol S(\Omega | \ell,m)$, which appear in MEG multipolar expansions \eqref{MPE-general}. In fact, Eq.~\eqref{MPE-general} follows by differentiation of Eq.~\eqref{MPE-potential} and restriction to sensor array locations $r=r(\Omega)$. Results relevant to our cases of interest are reported in Table \ref{TableA2}.
\begin{table}[ht]
\begin{center}
\setlength{\tabulinesep}{2pt} 
\begin{tabu}{|[1pt]c|[1.5pt]c|[1pt]}
    \tabucline[1pt]{1-2}
    \textbf{field component} & $S_i(\Omega | \ell,m)$ \\
    \tabucline[1.5pt]{1-2}
    $B_r=\frac{\partial U}{\partial r}$ & $-\frac{\ell+1}{r(\theta,\varphi)^{\ell+2}}Y_{\ell,m}(\theta,\varphi)$  \\
    \tabucline[1pt]{1-2}
    $B_\theta=\frac{1}{r}\frac{\partial U}{\partial \theta}$ & $\frac{1}{r(\theta,\varphi)^{\ell+2}}\frac{\partial Y_{\ell,m}(\theta,\varphi)}{\partial \theta}$  \\
        \tabucline[1pt]{1-2}
     $B_\varphi=\frac{1}{r \sin\theta}\frac{\partial U}{\partial \varphi}$ & $\frac{1}{r(\theta,\varphi)^{\ell+2} \sin\theta}\frac{\partial Y_{\ell,m}(\theta,\varphi)}{\partial \varphi}$ \\
    \tabucline[1pt]{1-2}
     $\frac{\partial B_r}{\partial r}$ & $\frac{(\ell+1)(\ell+2)}{r(\theta,\varphi)^{\ell+3}}Y_{\ell,m}(\theta,\varphi)$  \\
    \tabucline[1pt]{1-2}
      $\frac{1}{r}\frac{\partial B_r}{\partial \theta}$ & $-\frac{\ell+1}{r(\theta,\varphi)^{\ell+3}} \frac{\partial Y_{\ell,m}(\theta,\varphi)}{\partial \theta}$ \\
    \tabucline[1pt]{1-2}
    $\frac{1}{r \sin\theta}\frac{\partial B_r}{\partial \varphi}$ & $-\frac{\ell+1}{r(\theta,\varphi)^{\ell+3} \sin\theta} \frac{\partial Y_{\ell,m}(\theta,\varphi)}{\partial \varphi}$ \\
    \tabucline[1pt]{1-2}
    \end{tabu}
\end{center}
\caption{\label{TableA2}{{\it Elements of the vectorial spherical harmonics for each field or gradient component appearing in Table \ref{TableA1}.}} The angular location $\Omega$ is parameterised by the polar and azimuthal angles $(\theta,\varphi)$ depicted in Fig.~1 and the shape of the array, by $r=r(\theta,\varphi)$.}
\end{table}

\subsection{Results from the maximum-entropy hypothesis}\label{appendix-bckgd-MPEminor}

\paragraph{Data covariance} The significance of the maximum-entropy hypothesis \eqref{maxentropy} in our theory is to simplify the field covariance $\mathrm{cov}({\boldsymbol \phi})$ to $\sigma^2_{\boldsymbol a}\, \boldsymbol S\, \boldsymbol S^\dagger$. Along with the assumption \eqref{noisecov}, this implies that the $MN\times MN$ data covariance $\mathrm{cov}({\boldsymbol b})=\mathrm{cov}({\boldsymbol \phi})+\mathrm{cov}({\boldsymbol \varepsilon})$ associated with the observation model \eqref{obsmodel} reduces to
\begin{equation}\label{datcov} \mathrm{cov}({\boldsymbol b})=\sigma^2_{\boldsymbol a}\, \boldsymbol S\, \boldsymbol S^\dagger+\sigma_{\boldsymbol \varepsilon}^2\ \boldsymbol I\, ,
\end{equation}
with $\boldsymbol I$ denoting here the $NM\times NM$ identity matrix.
This relation has two consequences upon which our theory and numerical methods stand.

\paragraph{Multipole amplitude} First, plugging Eqs.~\eqref{datcov} and \eqref{noisecov} into our definition \eqref{snr-def} of the SNR demonstrates the relation \eqref{sigasigeps-est} that links the multipole amplitude parameter $\sigma_{\boldsymbol a}$ to experimental SNR measurements. This enables the determination of functional MEG multipolar expansions (Section \ref{Methods}).

\paragraph{Signal-space dimension} Second, Eq.~\eqref{datcov} provides a justification for our definition \eqref{sdof-a} of the signal-space dimension $\nu$. Each eigenvector of the matrix $\boldsymbol S\, \boldsymbol S^\dagger$ with eigenvalue $\lambda_u^2$ ($1\leq u\leq NM$) yields a neuromagnetic topography with a contribution $\sigma^2_{\boldsymbol a}\lambda_u^2$ to MEG signal variance. Comparison of this eigenvalue with the noise variance $\sigma_{\boldsymbol \varepsilon}^2$ (i.e., the unique eigenvalue of the noise covariance matrix, Eq.~\ref{noisecov}) determines whether this topography can be detected experimentally or is drowned  by noise. The number $\nu$ of detectable topographies, for which $ \sigma^2_{\boldsymbol a}\lambda_u^2$ exceeds $\sigma_{\boldsymbol \varepsilon}^2$, thus indeed corresponds to Eq.~\eqref{sdof-a}.

This definition of signal-space dimension turns out to be mathematically equivalent to the subtly different formulation
\begin{equation} \label{sdof-b} \nu = \#\left\{\, \textrm{eigenvalues $\lambda_u^2$ of $\boldsymbol S^\dagger \boldsymbol S\,$ with } \lambda_u^2 >  \sigma_{\boldsymbol \varepsilon}^2/\sigma^2_{\boldsymbol a} \, \right\} \, . \end{equation}
This is equivalent to Eq.~\eqref{sdof-a} because the $NM\times NM$ matrix $\boldsymbol S\, \boldsymbol S^\dagger$ and the infinite square matrix $\boldsymbol S^\dagger \boldsymbol S$ are hermitian conjugates of each other and thus share the same non-zero eigenvalues $\lambda_u^2$. The definition \eqref{sdof-a} is suitable for numerical evaluations (Section \ref{Methods}) whereas the reformulation \eqref{sdof-b} 
is better suited to the asymptotic analysis (\ref{appendix-proofs-spectrum}).

\subsection{A brief review of spherical mathematics}\label{appendix-bckgd-MPEreviewYlm}


\paragraph{Orthogonality relations on the sphere} Spherical harmonics play the same role for multipolar expansions \eqref{MPE-potential} than sines and cosines for Fourier expansions. They are solutions of the eigenvalue problem
\begin{subequations}\label{Ylm-def}
\begin{align} \label{Ylm-def-a} \boldsymbol \nabla^2 Y_{\ell,m} & = - \ell(\ell+1)\, Y_{\ell,m}\, ,  \\ \label{Ylm-def-b}  \frac{\partial Y_{\ell,m}}{\partial \varphi} & = \mathrm{i} m\, Y_{\ell,m}\, , \end{align}\end{subequations}
where $\boldsymbol \nabla^2$ denotes the laplacian operator on the unit sphere $\mathbb S$. The basic orthonormality property reads
\begin{equation}\label{Ylm-ON} \int_\mathbb{S} \mathrm d\Omega\, Y_{\ell,m}^* Y_{\ell',m'}^{} = \delta_{\ell,\ell'} \delta_{m,m'}\, , \end{equation}
with the $\delta$s referring to elements of the identity matrix. These integrals allow to demonstrate the orthogonality of vectorial spherical harmonics built from radial derivatives; this pertains to radial magnetometers and axial gradiometers (see Tables \ref{TableA1} and \ref{TableA2}). 

Vectorial spherical harmonics corresponding to planar sensors (magnetometers and gradiometers) include tangential gradients. Orthogonality holds too so long as polar and azimuthal components are combined in a rotationally invariant way, as is the case for planar magnetometers (Table \ref{TableA2}, second and third rows) and planar gradiometers (fifth and sixth rows). This may be proven on the basis of another, perhaps less standard, orthogonality property
\begin{equation}\label{Ylm-ONgrad} \int_\mathbb{S} \mathrm d\Omega\, \left[ \frac{\partial Y_{\ell,m}^*}{\partial \theta}\frac{\partial Y_{\ell',m'}^{}}{\partial \theta}+\frac{1}{(\sin \theta)^2} \frac{\partial Y_{\ell,m}^*}{\partial \varphi}\frac{\partial Y_{\ell',m'}^{}}{\partial \varphi} \right] = \ell (\ell+1)\, \delta_{\ell,\ell'} \delta_{m,m'}\, . \end{equation}
Since this relation is less standard than Eq.~\eqref{Ylm-ON}, we provide a quick demonstration. The integrand between brackets corresponds to the dot product $\boldsymbol \nabla Y_{\ell,m}^* \cdot \boldsymbol \nabla Y_{\ell',m'}^{}$ and can be replaced by $-Y_{\ell,m}^* \boldsymbol \nabla^2 Y_{\ell',m'}^{}$ after integration by parts; Eq.~\eqref{Ylm-ONgrad} then follows by direct application of the defining Eqs.~\eqref{Ylm-def-a} and \eqref{Ylm-ON}.

Neither Eq.~\eqref{Ylm-ON} nor Eq.~\eqref{Ylm-ONgrad} hold when restricting the closed, boundaryless sphere $\mathbb S$ to an open hemisphere. In the above argument, the boundary term generated by integration by parts would not vanish anymore and yield a cross-diagonal contribution (at least for certain values of the indices $\ell$ and $m$). The invariance of the integrand under local rotations (naturally implemented in single-axis radial sensors, but requiring summation of polar and azimuthal components in polar sensors) is also critical. Similar integrals for polar or azimuthal sensors separately would violate orthogonality.

\paragraph{Implementational details} We review the explicit representation of spherical harmonics that we implemented in our numerical analyses. We followed conventions widely used in electromagnetism and quantum physics \citep[see, e.g.,][]{Zangwill2012}. Specifically, we worked with
\begin{equation}\label{Ylm-rep} Y_{\ell,m}(\theta,\varphi) = \sqrt{\frac{2\ell+1}{4\pi}\frac{(\ell-|m|)!}{(\ell+|m|)!}}\, L_\ell^{|m|}(\mu) \, \mathrm{e}^{\mathrm{i} m \varphi}\, , \end{equation}
where we set $\mu=\cos\theta$ and the associated Legendre polynomials $L_\ell^m$ are defined for $m\geq 0$ by
\begin{equation}\label{Leg-rep} L_\ell^m(\mu) = (-)^m (1-\mu^2)^{m/2} \frac{\mathrm{d}^m}{\mathrm{d}\mu^m} \left[ \frac{1}{\ell!} \frac{\partial^\ell}{\partial x^\ell} \left(\frac{1}{\sqrt{1-2\mu x+x^2}}\right)\Bigg|_{x=0}\right]\, . \end{equation}
These expressions establish that $L_\ell^m(-\mu)=(-)^{\ell+m}\, L_\ell^m(\mu)$ and
\begin{equation}\label{Ylm-Z2} Y_{\ell,m}(\pi-\theta,\varphi)=(-)^{\ell+m}\, Y_{\ell,m}(\theta,\varphi)\, , \end{equation}
which embodies the (anti)symmetry of spherical harmonics upon exchanging the north and south hemispheres. This plays a key role for multipolar expansions on the hemiphere, as we describe below.

The representation \eqref{Ylm-rep} allowed to evaluate numerically the elements $S_i(\Omega|\ell,m)$ of the sensitivity matrix $\boldsymbol S$ corresponding to radial magnetometers and axial gradiometers (Table \ref{TableA2}). Polar derivatives relevant to planar sensors (Table \ref{TableA2}) were evaluated algebraically using
\begin{equation}\label{Ylm-rep_gradtheta} \frac{\partial Y_{\ell,m}}{\partial \theta} = - \sqrt{\frac{2\ell+1}{4\pi}\frac{(\ell-|m|)!}{(\ell+|m|)!}} \sin\theta \frac{\mathrm{d} L_\ell^{|m|}}{\mathrm{d}\mu} \, \mathrm{e}^{\mathrm{i} m \varphi} \end{equation}
and the recursive relations
\begin{equation}\label{Leg-recurse}  \frac{\mathrm{d} L_\ell^m}{\mathrm{d}\mu} =\begin{cases} -\frac{m\mu}{1-\mu^2} L_\ell^m-\frac{1}{\sqrt{1-\mu^2}}L_\ell^{m+1} & \textrm{for $0\leq m < \ell$,} \\ -\frac{m\mu}{1-\mu^2} L_\ell^m & \textrm{for $m=\ell$.} \end{cases} \end{equation}
The presence of (apparent) singularities at $\mu=\pm 1$ explains why we avoided the poles in our simulated spherical sensor grids (Section \ref{Methods}). Azimuthal derivatives were evaluated directly using the eigenvalue equation \eqref{Ylm-def-b}.

\subsection{Results on hemispherical multipolar expansions}\label{appendix-bckgd-MPEminorH}

\paragraph{Multipolar expansions on the hemisphere} The failure of general orthogonality properties on the hemisphere $\mathbb H$ is a sign that na\"ively restricting MEG multipolar expansion models to a hemispherical array is problematic. We show here how general expansions of the form \eqref{MPE-general} hold on $\mathbb H$ at the price of constraints enforced on the multipolar coefficients. Further, these constraints can, very fortunately, be chosen to ensure a version of the orthogonality properties \eqref{Ylm-ON} and \eqref{Ylm-ONgrad}. We outline the argument here since this situation is not standard.

Let us focus on the multipolar expansion \eqref{MPE-potential} of the magnetostatic potential $U=U(\theta,\varphi)$, setting $r=1$ to streamline notations (the dependence in $r$ can actually be absorbed into the multipolar coefficients $a_{\ell,m}$), so
\begin{equation}\label{MPE-potentialr1} U(\theta,\varphi)=\sum_{\ell,m} a_{\ell,m}\, Y_{\ell,m}(\theta,\varphi)\, . \end{equation}
Being able to apply this expansion requires knowing the potential $U$ on the whole sphere, but in the context of hemispherical MEG it is only measured over a hemisphere $\mathbb H$ that we will parameterise as the north hemisphere ($0\leq\theta\leq\pi/2$) for definiteness. The trick is to extend $U$ symmetrically to the south hemisphere ($\pi/2<\theta\leq \pi$) by setting $U(\theta,\varphi)=U(\pi-\theta,\varphi)$, which now renders Eq.~\eqref{MPE-potentialr1} legit. The symmetry of the extended potential upon exchanges of the north and south hemispheres constraints multipolar coefficients to $a_{\ell,m}=0$ whenever $\ell+m$ is odd due to the property \eqref{Ylm-Z2}. In other words, multipolar expansions on $\mathbb H$ are restricted to the subspace of symmetric spherical harmonics,
\begin{equation}\label{MPE-potentialr1H} U(\theta,\varphi)=\sum_\textrm{$\ell+m$ even} a_{\ell,m}\, Y_{\ell,m}(\theta,\varphi)\, . \end{equation}
This reduces the number of admissible values of $m$ from $2\ell+1$ on $\mathbb S$ to $\ell+1$ on $\mathbb H$. That is the mathematical origin of the difference between the MEG signal-space dimensions on the sphere \eqref{nu-sphere} and on the hemisphere \eqref{nu-hemisphere}.

\paragraph{Orthogonality relations on the hemisphere}
The symmetric spherical harmonics satisfy the orthogonality properties
\begin{equation}\label{Ylm-ONH} \int_\mathbb{H} \mathrm d\Omega\, Y_{\ell,m}^* Y_{\ell',m'}^{} = \tfrac{1}{2}\delta_{\ell,\ell'} \delta_{m,m'} \end{equation}
and
\begin{equation}\label{Ylm-ONgradH} \int_\mathbb{H} \mathrm d\Omega\, \left[ \frac{\partial Y_{\ell,m}^*}{\partial \theta}\frac{\partial Y_{\ell',m'}^{}}{\partial \theta}+\frac{1}{(\sin \theta)^2} \frac{\partial Y_{\ell,m}^*}{\partial \varphi}\frac{\partial Y_{\ell',m'}^{}}{\partial \varphi} \right] = \tfrac{\ell (\ell+1)}{2}\, \delta_{\ell,\ell'} \delta_{m,m'}\, . \end{equation}
These relations may be derived from their analog on the sphere (Eqs.~\ref{Ylm-ON} and \ref{Ylm-ONgrad}) and from the symbolic identity $\int_\mathbb{S}=2 \int_\mathbb{H}$, which is valid precisely because the integrands are all symmetric.

\section{The large-$N$ limit of MEG multipolar expansions: Asymptotic analysis}\label{appendix-proofs}
\setcounter{table}{0}
\setcounter{figure}{0}

We develop here the theory of multipolar expansions for MEG arrays covering a (hemi)sphere homogeneously and densely with a large number $N\rightarrow\infty$ of sensors. We consider first the asymptotics of the signal-space dimension $\nu$ and work towards the definition of the spatial resolution index $\ell_*$. Then, we derive our main estimation results (Eq.~\ref{lstar-subleading} and Eqs.~\ref{nu-sphere} and \ref{nu-hemisphere}).

\subsection{Asymptotics of signal-space dimension and definition of the spatial resolution index}\label{appendix-proofs-spectrum}

\paragraph{Continuous limit of $\boldsymbol S^\dagger \boldsymbol S$} The formulation \eqref{sdof-b} is more convenient for this large-$N$ analysis. Each entry $(\boldsymbol S^\dagger \boldsymbol S)_{\ell,m|\ell',m'}$ of the infinite matrix $\boldsymbol S^\dagger \boldsymbol S$ is a sum $ \sum_{\Omega} \boldsymbol S(\Omega | \ell,m)^\dagger\, \boldsymbol S(\Omega | \ell',m')$ running over the $N$ sensor locations $\Omega$. The dominant behaviour of this sum as $N\rightarrow\infty$
is controlled by the continuous integral\footnote{Equation \eqref{SS-continuous} follows merely from the definition of Riemann integrals, i.e., $\int \mathrm d\Omega \ldots = \lim_{N\rightarrow\infty} \sum_{\Omega} \delta\Omega \ldots$ with the element of solid angle corresponding to $ \delta\Omega=1/\rho $.
}
\begin{equation}\label{SS-continuous} \big(\boldsymbol S^\dagger \boldsymbol S\big)_{\ell,m|\ell',m'} \underset{N\rightarrow\infty}{\approx} \rho \int \mathrm d\Omega\,  \boldsymbol S(\Omega | \ell,m)^\dagger \boldsymbol S(\Omega | \ell',m')\, . \end{equation}
%
Integration runs here either over the whole sphere $\mathbb S$ for spherical MEG or over an hemisphere $\mathbb H$ for hemispherical MEG, and $\rho$ denotes the sensor density of the MEG array, i.e., the number of sensors per unit solid angle. For a homogeneous sensor coverage, we may set $\rho$ to either
\begin{equation}\label{SS-rho} \rho_\mathbb{S}=N/4\pi \quad \textrm{or} \quad \rho_\mathbb{H}=N/2\pi \end{equation}
depending on the case.
The right-hand side of Eq.~\eqref{SS-continuous} can be evaluated for each MEG system type defined in Table \ref{TableA1}, using for $\boldsymbol S(\Omega | \ell,m)$ the expressions given in Table \ref{TableA2} with constant, angle-independent $r$. All resulting integrals fall back to one of the orthogonality relations (Eqs.~\ref{Ylm-ON}, \ref{Ylm-ONgrad} for spherical MEG or Eqs.~\ref{Ylm-ONH}, \ref{Ylm-ONgradH} for hemispherical MEG). 
This allows to demonstrate that $\boldsymbol S^\dagger \boldsymbol S$ is asymptotically diagonal,
\begin{equation}\label{SS-diag}  \big(\boldsymbol S^\dagger \boldsymbol S\big)_{\ell,m|\ell',m'} \underset{N\rightarrow\infty}{\approx}\begin{cases}  \lambda_{\ell}^2 & \textrm{for $\ell=\ell'$ and $m=m'$,} \\ 0 & \textrm{for $\ell\neq\ell'$ or $m\neq m'$,} \end{cases} \end{equation}
with eigenvalues $\lambda_{\ell}^2$ listed in Table \ref{TableA3} (second column). Of note, these eigenvalues coincide numerically for spherical and hemispherical MEG because the extra halving factor appearing in Eqs.~\eqref{Ylm-ONH} and \eqref{Ylm-ONgradH} (compared to Eqs.~\ref{Ylm-ON} and \ref{Ylm-ONgrad}) cancels out exactly the doubling of sensor density indicated in Eq.~\eqref{SS-rho}, so $\rho_\mathbb{S}\int_\mathbb{S}\mathrm{d}\Omega\ldots=\rho_\mathbb{H}\int_\mathbb{H}\mathrm{d}\Omega\ldots$ applies.


\begin{table}[ht]
\begin{center}
\setlength{\tabulinesep}{2pt}
\begin{tabu}{|[1pt]c|[1.5pt]c|[1pt]c|[1pt]c|[1pt]c|[1pt]}
    \tabucline[1pt]{1-4}
    \textbf{sensor type} & $\lambda_\ell^2$ & $\mathsf P(\ell)$ & $\mathrm{deg}(\mathsf P)$  \\
    \tabucline[1.5pt]{1-4}
    radial magn & $\frac{N}{4\pi} \frac{(\ell+1)^2}{r^{2\ell+4}}$ & $(\ell+1)^2$ & $2$ \\
    \tabucline[1pt]{1-4}
    planar magn & $\frac{N}{4\pi} \frac{\ell(\ell+1)}{r^{2\ell+4}}$ & $\ell(\ell+1)$ & $2$ \\
    \tabucline[1pt]{1-4}
     tri-axis magn & $\frac{N}{4\pi} \frac{(\ell+1)(2\ell+1)}{r^{2\ell+4}}$ & $(\ell+1)(2\ell+1)$ & $2$ \\
    \tabucline[1pt]{1-4}
     axial grad & $\frac{N}{4\pi} \frac{(\ell+1)^2(\ell+2)^2}{r^{2\ell+6}}$ & $(\ell+1)^2(\ell+2)^2$ & $4$ \\
    \tabucline[1pt]{1-4}
     planar grad & $\frac{N}{4\pi} \frac{\ell(\ell+1)^3}{r^{2\ell+6}}$ & $\ell(\ell+1)^3$ & $4$ \\
    \tabucline[1pt]{1-4}

    \end{tabu}
\end{center}
\caption{\label{TableA3}{\it Large-$N$ eigenvalues of $\boldsymbol S^\dagger \boldsymbol S$ for (hemi)spherical MEG with sensor types defined in Table \ref{TableA1}.}  The modalitdependent polynomial $\mathsf P$ allows to gather all cases into the common Eq.~\eqref{SS-eig}. Its degree $\mathrm{deg}(\mathsf P)$ corresponds to twice the number of derivatives applied to the magnetostatic potential $U$ and distinguishes magnetometers from gradiometers
. The eigenvalues for Neuromag-type MEG arrays correspond to the sum of the eigenvalues for radial magnetometers and planar gradiometers. 
magn: magnetometer, grad: gradiometer.}
\end{table}

\paragraph{Scale-dependent MEG sensitivity profile} These eigenvalues measure the sensitivity of the MEG array to neuromagnetic field patterns at fixed angular frequency, i.e., any configuration $\boldsymbol \phi(\Omega)$ of the form $\sum_{m=-\ell}^\ell a_{\ell,m}\, \boldsymbol S(\Omega | \ell,m)$ with a fixed value of $\ell$. They are illustrated and discussed in a more pragmatic way in \ref{appendix-proofs-fig}. Continuing with formal developments, it proves convenient to summarize all cases described in Table \ref{TableA3} using the succinct expression
\begin{equation}\label{SS-eig} \lambda_{\ell}^2 = \frac{N}{4\pi} \frac{\mathsf P(\ell)}{r^{2\ell+2+\mathrm{deg}(\mathsf P)}} \,\cdot \end{equation}
%
The only factor varying across sensor types is the polynomial $\mathsf P$ controlling the precise dependence in the angular frequency and whose degree $\mathrm{deg}(\mathsf P)$ distinguishes between magnetometers and gradiometers (Table \ref{TableA3}). In fact, the general asymptotic eigenspectrum \eqref{SS-eig} works for any type of sensors (as long as they are combined in such a way that MEG recordings do not depend on sensor orientation along the sphere, so as to ensure orthogonality; see \ref{appendix-bckgd-MPEreviewYlm}). Results for multimodal setups mixing different sensor types (e.g., Neuromag-type arrays) are obtained by  adding the corresponding eigenvalues (since the matrix $\boldsymbol S^\dagger \boldsymbol S$ itself decomposes into a sum over sensor types, see Eq.~\ref{SS-continuous}). 
This additive behaviour may be readily verified using Table \ref{TableA3} in the case of
tri-axis magnetometers, which indeed corresponds to the combination of radial and planar magnetometers.

It is noteworthy that the large-$N$ eigenvalues \eqref{SS-eig} do not depend on the index $m$ so each $\lambda_{\ell}^2$ is degenerate, which is once again a consequence of the local rotation symmetry of MEG sensors. The level degeneracy is $2\ell +1$ for spherical MEG (as $m$ runs from $-\ell$ to $+\ell$ for multipolar expansions on $\mathbb S$) but it is reduced to $\ell+1$ for hemispherical MEG (as multipolar expansions on $\mathbb H$ are restricted to symmetric spherical harmonics with $\ell+m$ even; see Eq.~\ref{MPE-potentialr1H}). 

\paragraph{Spatial resolution index} Coming back to Eq.~\eqref{sdof-b}, the asymptotic diagonalisation result \eqref{SS-diag} shows that signal-space dimension may be computed as the number of eigenvalues $\lambda_\ell^2$ exceeding the threshold value $\sigma_{\boldsymbol \varepsilon}^2/\sigma^2_{\boldsymbol a}$, counting their degeneracy. Clearly, $\nu$ is controlled by the critical value of $\ell$ at the cross-over $\lambda_\ell^2=\sigma_{\boldsymbol \varepsilon}^2/\sigma^2_{\boldsymbol a}$, which corresponds precisely to the spatial resolution index $\ell_*$ outlined in the main text (Section \ref{Theory}). Of note, the criterion  works thanks to the degeneracy of large-$N$ eigenvalues; a \emph{bona fide} dependence in the index $m$ could lead to multiple values for $\ell_*$. Using Eq.~\eqref{SS-eig}, we thus find 
\begin{equation}\label{lstar-def} \frac{N}{4\pi} \frac{\mathsf P(\ell_*)}{r^{2\ell_*+2+\mathrm{deg}(\mathsf P)}} =  \frac{\sigma_{\boldsymbol \varepsilon}^2}{\sigma^2_{\boldsymbol a}} \, \cdot \end{equation}
Even though multipolar expansions \eqref{MPE-general} restrict the angular frequency index $\ell$ to integer values, the left-hand side of Eq.~\eqref{lstar-def} is an analytic function of $\ell_*$ and may thus be solved for real values of $\ell_*$.\footnote{The closest integer could be taken, but we shall not do so here. In any case, the difference between $\ell_*$ and its closest integer is well within the margins of error of Eq.~\eqref{SS-continuous}.} We derive the explicit solution and thus our main theoretical result \eqref{lstar-subleading} below.

\paragraph{Signal-space dimension} The argument developed above also allows to express the signal-space dimension $\nu$ in terms of $\ell_*$. For integer values of $\ell_*$, the large-$N$ estimate of $\nu$ can be obtained by summing all eigenvalue degeneracies up to $\ell=\ell_*$, so
\begin{equation}\label{nu-proof} \nu = \begin{cases} \nu_\mathbb S =\textstyle \sum_{\ell=0}^{\ell_*} (2\ell+1) = (\ell_*+1)^2\, , \\ \nu_\mathbb H =\textstyle \sum_{\ell=0}^{\ell_*} (\ell+1) = \tfrac{1}{2} (\ell_*+1)(\ell_*+2)\, , \end{cases} \end{equation}
%
%
for spherical and hemispherical MEG, respectively. The right-hand sides then allow to extend the notion of signal-space dimension to any non-integer solution of Eq.~\eqref{lstar-def}. This completes the proof of Eqs.~\eqref{nu-sphere} and \eqref{nu-hemisphere}.

%
\begin{figure*}\centering
\includegraphics[scale=0.9]{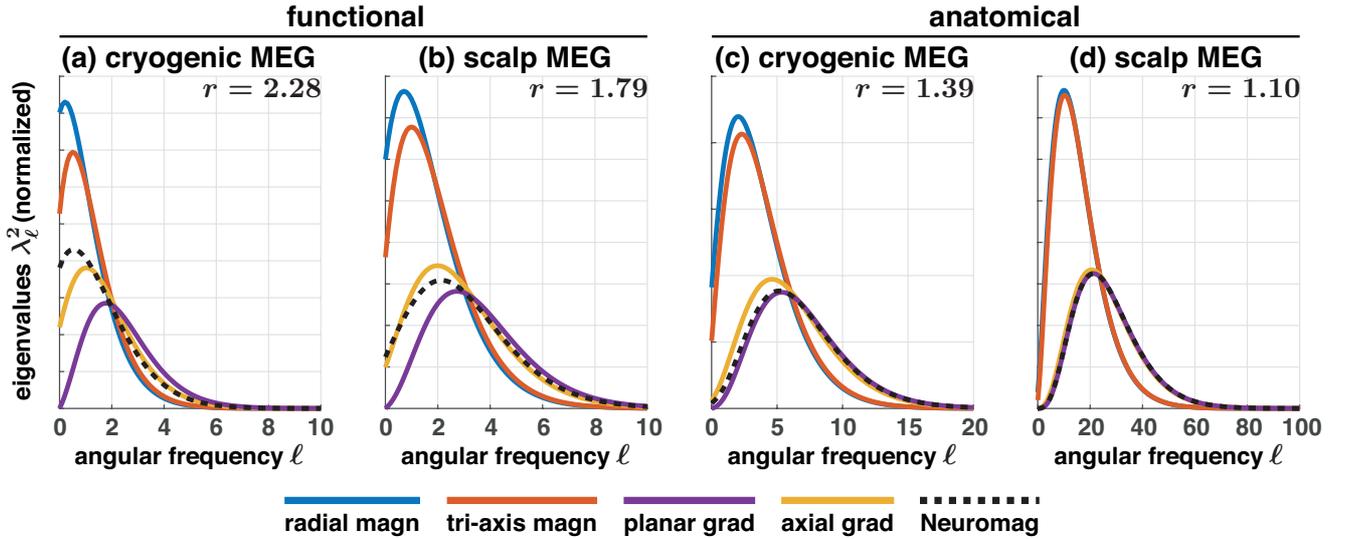}
\caption{\label{figA} \emph{Scale-dependent sensitivity of (hemi)spherical MEG arrays in the large-$N$ limit.} The spectrum of large-$N$ eigenvalues $\lambda_\ell^2$ is plotted as a function of the angular frequency index $\ell$ for several MEG sensor types and for sensor-to-brain distances set using Table \ref{Table4} to functional (\textbf{left}) or anatomical (\textbf{right}) estimates corresponding to cryogenic (panels \textbf{a}, \textbf{c}) or scalp MEG (panels \textbf{b}, \textbf{d}). Each curve was drawn based on Table \ref{TableA3} with further normalization to a unit area under the curve for visualization purposes; the lower peak amplitudes for gradiometers thus correspond to larger relative contributions of high angular frequencies than magnetometers. The case of the Neuromag-type array mixing radial magnetometers and planar gradiometers was obtained by summing the corresponding eigenvalues (Table \ref{TableA3}). magn: magnetometer, grad: gradiometer.}
\end{figure*}

\subsection{Scale-dependent sensitivity profile of large-$N$ MEG arrays}\label{appendix-proofs-fig}
Let us make a pause in our mathematical analysis and illustrate in Figure \ref{figA} the $\ell$-dependent profile of the large-$N$ eigenvalues \eqref{SS-eig}, for different (hemi)spherical MEG arrays at the four sensor-to-brain distances reported in the main text (Table \ref{Table4}). These spectra provide a measure of MEG sensitivity across spatial scales indexed by the angular frequency index $\ell$ (low frequencies correspond to broad topographies and high frequencies, to focal topographies).

Magnetometers exhibited sensitivity profiles peaking at lower angular frequency, and with lesser relative contribution from high angular frequencies, than gradiometers. This reflects the better ability of magnetometers to probe widespread neuromagnetic topographies (corresponding to deep brain activity), and that of gradiometers to probe focal topographies (cortical activity). In functional MEG inferred from resting-sate recordings (Fig.~\ref{figA}a,b), magnetometric sensitivity appeared optimal for homogeneous (angular frequency $\ell=0$) and dipolar ($\ell=1$) fields, with tri-axis sensors receiving a slightly larger relative contribution from more focal fields than radial sensors alone. Optimum gradiometric sensitivity varied from dipolar ($\ell=1$, cryogenic axial gradiometers corresponding to a CTF-like MEG) to octopolar fields ($\ell=3$, scalp planar gradiometers). Interestingly, cryogenic CTF-like MEG strikes a balance between magnetometers and planar gradiometers; a similar balance is achieved in Neuromag-like MEG by combining magnetometers and planar gradiometers. 
It is noteworthy that both MEG systems would converge if they were placed directly on scalp (Fig.~\ref{figA}b).

Further moving sensors towards the scalp increased sensitivity to focal topographies, as expected and clearly illustrated with anatomical MEG (Fig.~\ref{figA}c,d; compare abscissa scales across panels a--d). A striking observation from Fig.~\ref{figA}d is the disappearence of differences between radial and tri-axis magnetometers, and between axial and planar gradiometers. Sensitivity profiles converged into one of two distinct classes as sensors approach the scalp; one gathering all magnetometric MEG and the other, all gradiometric MEG (including the Neuromag-like system). The lack of impact of sensor type is derived more formally in \ref{appendix-proofs-properties}.


\subsection{Large-$N$ analysis of Eq.~\eqref{lstar-def}}\label{appendix-proofs-largeNsoln}

\paragraph{Leading behaviour} We now solve Eq.~\eqref{lstar-def} explicitly in the large-$N$ limit and derive our main theoretical result, i.e., Eq.~\eqref{lstar-subleading}. For that purpose, it is mathematically convenient to take the logarithm of Eq.~\eqref{lstar-def} and analyze
\begin{equation}\label{lstar-def-log}  2 \ell_* \log r - \log\mathsf P(\ell_*) = \log \left(  \frac{N}{4\pi\, r^{2+\mathrm{deg}(\mathsf P)}} \frac{\sigma_{\boldsymbol a}^2}{\sigma_{\boldsymbol \varepsilon}^2} \right) \, . \end{equation}
The part of the right-hand side that dominates as $N\rightarrow\infty$ is $\log N$. In turn, the left-hand side must diverge, so $\ell_*$ is large and dominates over the second term $\log \mathsf P(\ell_*)$.\footnote{The alternative would be that $\ell_*$ approaches a zero of the polynomial $\mathsf P$, since then $\log\mathsf P(\ell_*)\rightarrow -\infty$. Examination of Table \ref{TableA3} (third column) reveals that all such solutions are non-positive and thus physically unacceptable for our purposes.} The  leading behaviour of the solution to Eq.~\eqref{lstar-def-log} is thus controlled by the linear equation $2\ell_* \log r = \log N$, so that
\begin{equation}\label{lstar-leading} \ell_* = \frac{\log N}{2\log r} + \textrm{sub-leading corrections.} \end{equation}

\paragraph{Sub-leading corrections} To proceed further, it proves convenient to introduce some notations. Parameterising the modality-dependent polynomial $\mathsf P$ (Table \ref{TableA3}) as
\begin{equation}\label{P-largel} \mathsf P(\ell)=a \ell^b \left[1+\mathcal O(1/\ell)\right] \end{equation}
for large $\ell$, straigthforward algebraic manipulations allow to recast Eq.~\eqref{lstar-def-log} in the condensed form
\begin{equation}\label{lstar-recast} \ell_* - k \log\ell_* + \mathcal O(1/\ell_*) = \Lambda\, , \end{equation}
where we defined for convenience
\begin{equation}\label{k-Lambda} k = \frac{b}{2 \log r} \quad \textrm{and} \quad \Lambda = \frac{\log \left(  \frac{a\, N}{4\pi\, r^{2+b}} \frac{\sigma_{\boldsymbol a}^2}{\sigma_{\boldsymbol \varepsilon}^2} \right)}{2\log r}\, \cdot \end{equation}
The latter parameter diverges logarithmically as $N\rightarrow\infty$, so we seek to solve Eq.~\eqref{lstar-recast} in the limit $\Lambda\rightarrow\infty$. For future reference, we note that
\begin{equation}\label{logLambda} \log\Lambda = \log\left[\frac{\log N}{2\log r}\right]+\mathcal O\left(1/\log N\right)\, . \end{equation}

Our intermediate result \eqref{lstar-leading} suggests that $\ell_*\approx \Lambda$. Using a change of variable $\ell_* = \Lambda \left(1+\varepsilon\right)$ emphasizing the relative correction $\varepsilon$ to this leading behaviour, Eq.~\eqref{lstar-recast} becomes
\begin{equation}\label{epsilon-recast} \varepsilon -  \frac{k \log\left(1+\varepsilon\right)}{\Lambda} = \frac{k \log\Lambda}{\Lambda} + \mathcal O(1/\Lambda^2)\, .  \end{equation}
Taking the limit $\Lambda\rightarrow\infty$ implies that $\varepsilon\rightarrow0$, and the solution to leading order is $\varepsilon = \frac{k \log\Lambda}{\Lambda}+\mathcal O\left(\frac{\log\Lambda}{\Lambda^2}\right)$ or equivalently
\begin{equation} \ell_* = \Lambda + k \log\Lambda+\mathcal O\left(\frac{\log\Lambda}{\Lambda}\right)\, . \end{equation}
%

Going back to the original parameterisation using Eqs.~(\ref{k-Lambda}--\ref{logLambda}) and rearranging terms, we obtain at last
\begin{equation}\label{lstar-subleadingA} \ell_* = \frac{\log \left[ \frac{N}{4\pi\, r^{2+d}} \frac{\sigma_{\boldsymbol a}^2}{\sigma_{\boldsymbol \varepsilon}^2}\, a \left(\frac{\log N}{2\log r}\right)^b \right]}{2\log r} + \mathcal O\left(\frac{\log\log N}{\log N}\right)\, . \end{equation}
This corresponds to our main theoretical result \eqref{lstar-subleading}, if the factor $a\left( \log N / 2 \log r\right)^b$ may be replaced by $\mathsf P\left(\log N / 2 \log r\right)$. This replacement is legit up a vanishingly small error of order $1/\log N$ (see Eq.~\ref{P-largel} with $\ell=\log N / 2 \log r$), which is subdominant compared to the error term in Eq.~\eqref{lstar-subleadingA}. That completes the proof of Eq.~\eqref{lstar-subleading}.

\subsection{Derivation of several properties of high-density MEG spatial resolution from Eq.~\eqref{lstar-subleading}}\label{appendix-proofs-properties}

\paragraph{Logarithmic slowness of spatial resolution gains and the cost of spatial resolution} The leading behaviour \eqref{lstar-leading} explicits the logarithmic divergence of MEG spatial resolution as $N\rightarrow\infty$ (property \ref{item-density} of main text Section \ref{Theory}). The slowness of this divergence may be interpreted by considering the gain $\Delta\ell_*$ of spatial resolution index obtained by adding one sensor (i.e., increasing $N$ to $N+1$). Equation \eqref{lstar-leading} shows that $\Delta\ell_*\approx\log\tfrac{N+1}{N}\approx 1/N$, which is negligibly small at large $N$. Improving spatial resolution by adding sensors to an already high-density MEG array becomes increasingly difficult.

Evaluating the cost of spatial resolution provides an alternative, more pragmatic view of this difficulty (Section \ref{Discussion}). We measure here this cost as the number $\Delta N$ of additional sensors needed to increase the spatial resolution index $\ell_*$ by a small, fixed amount of $\Delta \ell_*=1$. This can be estimated as the inverse of the slope $\partial \ell_* / \partial N$. In the high-density regime at very large $N$,
Eq.~\eqref{lstar-subleading} leads to a cost $\Delta N\approx 2 \log r\times N$ that increases linearly with $N$. By contrast, in the low-density regime described by Eq.~\eqref{lstar-linear}, $\Delta N=2\sqrt{N/M}$ so the costs for improving spatial resolution are much lower at sufficiently small $N$.

\paragraph{Near-brain behaviour, sub-leading divergence, and the impact of magnetometers/gradiometers} Equation \eqref{lstar-leading} also establishes our claim that the sensor-to-brain distance $r$ is the only parameter that modulates this leading behaviour, and that dependence includes a divergence $1/\log r\approx (r-1)^{-1}$ as $r\rightarrow 1$ (Section \ref{Theory}, property \ref{item-radius}).

This divergence also affects the sub-leading corrections in Eq.~\eqref{lstar-leading}. In the limit $r\rightarrow 1$, the full solution \eqref{lstar-subleading} behaves as 
\begin{equation}\label{lstar-nearbrain} \ell_*\underset{r\rightarrow 1}{\approx}\mathrm{deg}(\mathsf P)\times \frac{\log\left(\log N / \log r\right)}{2\log r}\, \cdot \end{equation}
Table \ref{TableA3} reveals that the polynomial degree $\mathrm{deg}(\mathsf P)$ distinguishes magnetometers and gradiometers (with a value twice larger for the latter) but not the number of recorded components or their orientation.
This establishes our claim that magnetometers and gradiometers are discriminated at the level of a sub-leading divergence in $N$ (Section \ref{Theory}, property \ref{item-type}), which turns out to be doubly logarithmic.

\paragraph{Insensitivity to sensor components, SNR, and sensor coverage} Equations \eqref{lstar-leading} and \eqref{lstar-nearbrain} disclose the only two terms of the full solution \eqref{lstar-subleading} that diverge as $N\rightarrow\infty$; all other terms are either finite or fall off with $N$. This establishes our claims that high-density MEG spatial resolution is essentially insensitive to the number and orientation of sensor components (Section \ref{Theory}, property \ref{item-type}) and to the exact level of SNR (Section \ref{Theory}, property \ref{item-snr}). For example, expanding the logarithm in Eq.~\eqref{lstar-subleading} shows that the SNR only contributes through the finite, subtle correction $\log(\sigma_{\boldsymbol a}/\sigma_{\boldsymbol \varepsilon})/\log r$.

This quasi-insensitivity to the SNR is a significant aspect of our theory. It shows that our measurement of MEG spatial resolution based on Eq.~\eqref{lstar-subleading} does not depend on the precise choice of noise level used to define signal-space dimension \eqref{sdof-a} and the spatial resolution index \eqref{lstar-def}. Other authors might choose an eigenvalue threshold different than ours, say $\eta\times\sigma_{\boldsymbol \varepsilon}^2/\sigma^2_{\boldsymbol a}$ with $\eta$ of order one but different from one; however, the impact of choosing $\eta\neq 1$ on the spatial resolution index would be restricted to the addition of a finite, negligible correction $\log \eta / 2 \log r$ in the right-hand size of Eq.~\eqref{lstar-subleading}.

We also demonstrated that spatial resolution is independent of sensor coverage (see discussion below Eq.~\ref{SS-diag}). On the other hand, that is not true for the signal-space dimension $\nu$ since it appears to scale with sensor coverage as $\nu_\mathbb H \approx \nu_\mathbb S / 2$ (Section \ref{Theory} and \ref{appendix-proofs-spectrum}). A more precise comparison of Eqs.~\eqref{nu-sphere} and \eqref{nu-hemisphere} leads to
%
\begin{equation} \nu_\mathbb{H}=\frac{\nu_\mathbb{S}}{2}\left(1+\frac{1}{\sqrt{\nu_\mathbb{S}}}\right) > \frac{\nu_\mathbb{S}}{2}\, \cdot \end{equation}
This establishes formally our claim that the ``field of view'' of hemispherical MEG extends beyond the borders of its hemispherical array (Section \ref{Discussion}); though this extension is fairly limited at high spatial resolution since $1/\sqrt{\nu_\mathbb S} \approx 1/\ell_*$ (see Eq.~\ref{nu-sphere}) is small at large $N$.

\paragraph{Spatial degrees of freedom per sensor and oversampling} The leading behaviour \eqref{lstar-leading} for $\ell_*$ combined with Eq.~\eqref{nu-hemisphere} for $\nu$ allows to show that the number of independent degrees of freedom available per sensor
\begin{equation}\label{nu-oversampling} \nu_\mathbb H/N \underset{N\rightarrow \infty}{\approx} \frac{(\log N)^2}{8 N (\log r)^2} \end{equation}
is negligibly small at large $N$. This is again a reflection of the difficulty to harvest extra information by adding sensors to already high-density MEG arrays.

This property is closely related to the notion of oversampling, whose formal definition may be written as $MN/\nu\gg 1$. From Eq.~\eqref{nu-oversampling}, we find $MN/\nu \approx 8MN (\log r/\log N)^2$, so the condition is fullfiled at large $N$. This confirms formally that MEG is oversampled in the high-density regime (Section \ref{Discussion}). Note also how increasing both the number $M$ of sensor components and the sensor-to-brain distance $r$ further help reaching oversampling. By contrast, $MN/\nu = 1$ in the low-density regime described by Eq.~\eqref{nu-linear}.

\section{On the physics of the low/high-density transition and the size of magnetic field smoothness}\label{appendix-smoothness}
\setcounter{table}{0}
\setcounter{figure}{0}

We discuss here how to take advantage of our description of MEG spatial resolution at the transition between the low- and high-density regimes to assess the typical size of magnetic field smoothness. We use this analysis to partly justify our description of the spatial resolution index in the low-density regime (Eq.~\ref{lstar-linear}) and to elucidate the fairly counter-intuitive physical mechanism underlying how MEG spatial resolution increases as sensors are brought closer to sources of brain activity (main text, Section \ref{Discussion}).

\paragraph{Scale of magnetic field smoothness}
We seek to estimate the size $\lambda$ of magnetic field smoothness. 
Physically, what should control MEG spatial resolution is the inter-sensor separation $s$ relative to field smoothness size $\lambda$. When $s\gg \lambda$, sensors are insensitive to field smoothness and they bring independent information; this is the low-density regime (Eq.~\ref{nu-linear}). When $s\ll \lambda$, field smoothness strongly limits the amount of independent information and constraints the limits of spatial resolution; this is the high-density regime (Eq.~\ref{lstar-leading} to leading order). The in-between situation $s\approx \lambda$ corresponds to the transition.

We conclude that field spread size $\lambda$ can be estimated as the inter-sensor separation $s$ at the transition. On the one hand, we have
\begin{equation}\label{sensordist} s\approx \sqrt{\frac{8}{N}}\, R_\textrm{array}\, , \end{equation}
at least when $N$ is large enough. This follows from the geometric condition that the surface available for each sensor ($\textrm{area}=2\pi R_\textrm{array}^2/N$) corresponds to a small spherical cap of diameter $s$ surrounding the sensor ($\textrm{area}\approx\pi s^2/4$). On the other hand, our results allow to describe the transition line (Fig.~\ref{fig4}) parametrically as $N=N_t(r)$. To leading order, we have
\begin{equation}\label{transition-param} \frac{\log N_t}{\sqrt{N_t}}\approx 2\sqrt{M} \log r \end{equation}
with $M$ the number of sensor components and $r=R_\textrm{array}/R_\textrm{brain}$. This follows from the condition that the low-density ansatz ($\ell_*\approx \sqrt{MN}$ to leading order; see Eq.~\ref{lstar-linear}) and the high-density prediction (leading order in Eq.~\ref{lstar-leading}) coincide. Taken together, these two results yield the estimate
%
\begin{equation}\label{fieldspread-size} \lambda \approx  \sqrt{\frac{8}{N_t(r)}} \, R_\textrm{array}\, . \end{equation}
We restricted our description \eqref{transition-param} of the low/high-density transition to the leading order for simplicity, so our result \eqref{fieldspread-size} provides a rough estimate rather than a \vince{numerically accurate} estimate. We can nevertheless apply it to answer two qualitative questions that appeared in our analysis and discussion.



\paragraph{Consistency of the low-density ansatz (Eq.~\ref{lstar-linear})} On physical grounds, magnetic field smoothness acts on neuromagnetic topographies as a filter that lets pass spatial scales below its size $\lambda$ (Section \ref{Discussion}). It does not affect neuromagnetic recordings in the low-density regime $s\gg\lambda$ (where spatial resolution is then controlled by the total number of recordings; that is the content of Eq.~\ref{lstar-linear}) and starts to take effect at the transition ($N\approx N_t$). At this point, the order of magnitude of field smoothness size ($R_\textrm{array}/\sqrt{N}$) must coincide with the order of magnitude of the largest wavelength accessible to the MEG array ($R_\textrm{array}/\sqrt{\ell_*(\ell_*+1)}\approx R_\textrm{array}/\ell_*$). We conclude that the spatial resolution index $\ell_*$ scales with $\sqrt{N}$ along the transition line $N=N_t(r)$. This square-root law is fully consistent with our ansatz \eqref{lstar-linear} describing spatial resolution at the transition between the low- and high-density regimes.

\paragraph{Effect of reducing the sensor-to-brain distance} We discussed in Section \ref{Discussion} that shrinking a MEG array reduces the inter-source distance relative to field smoothness size. This claim may be established formally by considering the ratio of Eqs.~\eqref{sensordist} and \eqref{fieldspread-size}, i.e.,
\begin{equation} s/\lambda \approx \sqrt{N_t(r)/N}\, . \end{equation}
Our parametric description of $N_t(r)$ shows that it increases when the sensor-to-brain distance $r$ decreases; e.g., taking the derivative of Eq.~\eqref{transition-param} with respect to $r$ yields
\begin{equation} \frac{\mathrm d N_t}{\mathrm d r} \approx \frac{4 \sqrt{M N_t^3}}{2-\log N_t} \end{equation}
which is negative to leading order $N_t\gg 1$. We conclude that shrinking a MEG array (i.e., reducing $R_\textrm{array}$ with $N$ kept fixed) augments the inter-source distance measured in units of field smoothness size and thus reduces sensor density measured in units of field smoothness area, even though it has the opposite effect on geometric separation \eqref{sensordist} and sensor density ($N/2\pi R_\textrm{array}^2$).


\section*{References}

\bibliographystyle{elsarticle-harv}
\bibliography{./references}

\begin{thebibliography}{54}
\expandafter\ifx\csname natexlab\endcsname\relax\def\natexlab#1{#1}\fi
\expandafter\ifx\csname url\endcsname\relax
  \def\url#1{\texttt{#1}}\fi
\expandafter\ifx\csname urlprefix\endcsname\relax\def\urlprefix{URL }\fi

\bibitem[{Ahonen et~al.(1993)Ahonen, H{\"a}m{\"a}l{\"a}inen, Ilmoniemi, Kajola,
  Knuutila, Simola, and Vilkman}]{Ahonen1993}
Ahonen, A., H{\"a}m{\"a}l{\"a}inen, M., Ilmoniemi, R., Kajola, M., Knuutila,
  J., Simola, J., Vilkman, V., 1993. Sampling theory for neuromagnetic detector
  arrays. IEEE Tran. Biomed. Eng. 40~(9), 859--869.

\bibitem[{Barnes et~al.(2011)Barnes, Litvak, Brookes, and Friston}]{Barnes2011}
Barnes, G.~R., Litvak, V., Brookes, M.~J., Friston, K.~J., 2011. Controlling
  false positive rates in mass-multivariate tests for electromagnetic
  responses. NeuroImage 56~(3), 1072—1081.

\bibitem[{Beltrachini et~al.(2021)Beltrachini, von Ellenrieder, Eichardt, and
  Haueisen}]{Beltrachini2021}
Beltrachini, L., von Ellenrieder, N., Eichardt, R., Haueisen, J., 2021. Optimal
  design of on-scalp electromagnetic sensor arrays for brain source
  localisation. Human Brain Mapping 42~(15), 4869--4879.

\bibitem[{Borna et~al.(2020)Borna, Carter, Colombo, Jau, McKay, Weisend, Taulu,
  Stephen, and Schwindt}]{Borna2020}
Borna, A., Carter, T., Colombo, A., Jau, Y., McKay, J., Weisend, M., Taulu, S.,
  Stephen, J., Schwindt, P., 2020. Non-invasive functional-brain-imaging with
  an {OPM}-based magnetoencephalography system. PLoS One 15, e0227684.

\bibitem[{Boto et~al.(2016)Boto, Bowtell, Krüger, Fromhold, Morris, Meyer,
  Barnes, and Brookes}]{Boto2016}
Boto, E., Bowtell, R., Krüger, P., Fromhold, T., Morris, P., Meyer, S.,
  Barnes, G., Brookes, M., 2016. {On the potential of a new generation of
  magnetometers for MEG: A Beamformer simulation study}. PLoS One 11, e0157655.

\bibitem[{Boto et~al.(2021)Boto, Hill, Rea, Holmes, Seedat, Leggett, Shah,
  Osborne, Bowtell, and Brookes}]{Boto2021}
Boto, E., Hill, R.~M., Rea, M., Holmes, N., Seedat, Z.~A., Leggett, J., Shah,
  V., Osborne, J., Bowtell, R., Brookes, M.~J., 2021. Measuring functional
  connectivity with wearable {MEG}. NeuroImage 230, 117815.

\bibitem[{Boto et~al.(2018)Boto, Holmes, Leggett, Roberts, Shah, Meyer,
  Duque~Muñoz, Mullinger, Tierney, Bestmann, Barnes, Bowtell, and
  Brookes}]{Boto2018}
Boto, E., Holmes, N., Leggett, J., Roberts, G., Shah, V., Meyer, S.~S.,
  Duque~Muñoz, L., Mullinger, K.~J., Tierney, T.~M., Bestmann, S., Barnes,
  G.~R., Bowtell, R., Brookes, M.~J., 2018. {Moving magnetoencephalography
  towards real-world applications with a wearable system}. Nature 555,
  657--661.

\bibitem[{Brookes et~al.(2021)Brookes, Boto, Rea, Shah, Osborne, Holmes, Hill,
  Leggett, Rhodes, and Bowtell}]{Brookes2021}
Brookes, M.~J., Boto, E., Rea, M., Shah, V., Osborne, J., Holmes, N., Hill,
  R.~M., Leggett, J., Rhodes, N., Bowtell, R., 2021. Theoretical advantages of
  a triaxial optically pumped magnetometer magnetoencephalography system.
  NeuroImage 236, 118025.

\bibitem[{Coquelet et~al.(2020)Coquelet, {De Ti\`ege}, Destoky, Roshchupkina,
  Bourguignon, Goldman, Peigneux, and Wens}]{Coquelet2020}
Coquelet, N., {De Ti\`ege}, X., Destoky, F., Roshchupkina, L., Bourguignon, M.,
  Goldman, S., Peigneux, P., Wens, V., 2020. {Comparing MEG and high-density
  EEG for intrinsic functional connectivity mapping}. NeuroImage 210, 116556.

\bibitem[{Coquelet et~al.(2022)Coquelet, {De Ti\`ege}, Roshchupkina, Peigneux,
  Goldman, Woolrich, and Wens}]{Coquelet2022}
Coquelet, N., {De Ti\`ege}, X., Roshchupkina, L., Peigneux, P., Goldman, S.,
  Woolrich, M., Wens, V., 2022. {Microstates and power envelope hidden Markov
  modeling probe bursting brain activity at different timescales}. NeuroImage
  247, 118850.

\bibitem[{Dale and Sereno(1993)}]{Dale1993}
Dale, A.~M., Sereno, M.~I., 1993. {Improved localization of cortical activity
  by combining EEG and MEG with MRI cortical surface reconstruction: A linear
  approach}. J Cogn Neurosci 5~(2), 162--176.

\bibitem[{Duque-Muñoz et~al.(2019)Duque-Muñoz, Tierney, Meyer, Boto, Holmes,
  Roberts, Leggett, Vargas-Bonilla, Bowtell, Brookes, López, and
  Barnes}]{Duque-Munoz2019}
Duque-Muñoz, L., Tierney, T.~M., Meyer, S.~S., Boto, E., Holmes, N., Roberts,
  G., Leggett, J., Vargas-Bonilla, J.~F., Bowtell, R., Brookes, M.~J., López,
  J.~D., Barnes, G.~R., 2019. Data-driven model optimization for optically
  pumped magnetometer sensor arrays. Human Brain Mapping 40~(15), 4357--4369.

\bibitem[{Feys et~al.(2022)Feys, Corvilain, Aeby, Sculier, Holmes, Brookes,
  Goldman, Wens, and De~Ti\`{e}ge}]{Feys2022}
Feys, O., Corvilain, P., Aeby, A., Sculier, C., Holmes, N., Brookes, M.,
  Goldman, S., Wens, V., De~Ti\`{e}ge, X., 2022. On-scalp optically pumped
  magnetometers versus cryogenic magnetoencephalography for diagnostic
  evaluation of epilepsy in school-aged children. Radiology 0~(0), 212453.

\bibitem[{Fischl(2012)}]{Fischl2012}
Fischl, B., 2012. {FreeSurfer}. Neuroimage 62, 774--781.

\bibitem[{Gramfort et~al.(2014)Gramfort, Luessi, Larson, Engemann, Strohmeier,
  Brodbeck, Parkkonen, and H{\"a}m{\"a}l{\"a}inen}]{Gramfort2014}
Gramfort, A., Luessi, M., Larson, E., Engemann, D.~A., Strohmeier, D.,
  Brodbeck, C., Parkkonen, L., H{\"a}m{\"a}l{\"a}inen, M.~S., 2014. {MNE
  software for processing MEG and EEG data}. Neuroimage 86, 446--460.

\bibitem[{H\"am\"al\"ainen et~al.(1993)H\"am\"al\"ainen, Hari, Ilmoniemi,
  Knuutila, and Lounasmaa}]{Hamalainen1993MEG}
H\"am\"al\"ainen, M., Hari, R., Ilmoniemi, R.~J., Knuutila, J., Lounasmaa,
  O.~V., 1993. Magnetoencephalography---theory, instrumentation, and
  applications to noninvasive studies of the working human brain. Rev Mod Phys
  65, 413--497.

\bibitem[{Hari and Puce(2017)}]{HariPuce}
Hari, R., Puce, A., 2017. {MEG-EEG primer; 1st ed.} Oxford University Press,
  Oxford, UK.

\bibitem[{Hauk and Stenroos(2014)}]{Hauk2014}
Hauk, O., Stenroos, M., 2014. {A framework for the design of flexible
  cross-talk functions for spatial filtering of EEG/MEG data: DeFleCT}. Hum
  Brain Mapp 35~(4), 1642--1653.

\bibitem[{Hill(1954)}]{Hill}
Hill, E.~L., 1954. The theory of vector spherical harmonics. Amer J Phys
  22~(4), 211--214.

\bibitem[{Hill et~al.(2020)Hill, Boto, Rea, Holmes, Leggett, Coles,
  Papastavrou, Everton, Hunt, Sims, Osborne, Shah, Bowtell, and
  Brookes}]{Hill2020}
Hill, R.~M., Boto, E., Rea, M., Holmes, N., Leggett, J., Coles, L.~A.,
  Papastavrou, M., Everton, S.~K., Hunt, B.~A., Sims, D., Osborne, J., Shah,
  V., Bowtell, R., Brookes, M.~J., 2020. Multi-channel whole-head {OPM-MEG}:
  Helmet design and a comparison with a conventional system. NeuroImage 219,
  116995.

\bibitem[{Holmes et~al.(2018)Holmes, Leggett, Boto, Roberts, Hill, Tierney,
  Shah, Barnes, Brookes, and Bowtell}]{Holmes2018}
Holmes, N., Leggett, J., Boto, E., Roberts, G., Hill, R.~M., Tierney, T.~M.,
  Shah, V., Barnes, G.~R., Brookes, M.~J., Bowtell, R., 2018. A bi-planar coil
  system for nulling background magnetic fields in scalp mounted
  magnetoencephalography. NeuroImage 181, 760--774.

\bibitem[{Holmes et~al.(2019)Holmes, Tierney, Leggett, Boto, Mellor, Roberts,
  Hill, Shah, Barnes, Brookes, and Bowtell}]{Holmes2019}
Holmes, N., Tierney, T., Leggett, J., Boto, E., Mellor, S., Roberts, G., Hill,
  R., Shah, V., Barnes, G., Brookes, M., Bowtell, R., 2019. Balanced, bi-planar
  magnetic field and field gradient coils for field compensation in wearable
  magnetoencephalography. Sci. Rep. 9, 14196.

\bibitem[{Hyv{\"a}rinen and Oja(2000)}]{HyvarinenOja2000}
Hyv{\"a}rinen, A., Oja, E., 2000. Independent component analysis: algorithms
  and applications. Neural Netw 13, 411--430.

\bibitem[{Iivanainen et~al.(2021)Iivanainen, M\"akinen, Zetter, Stenroos,
  Ilmoniemi, and Parkkonen}]{Iivanainen2021}
Iivanainen, J., M\"akinen, A.~J., Zetter, R., Stenroos, M., Ilmoniemi, R.~J.,
  Parkkonen, L., 2021. {Spatial sampling of MEG and EEG based on generalized
  spatial-frequency analysis and optimal design}. NeuroImage 245, 118747.

\bibitem[{Iivanainen et~al.(2017)Iivanainen, Stenroos, and
  Parkkonen}]{Ilvanainen2017}
Iivanainen, J., Stenroos, M., Parkkonen, L., 2017. Measuring {MEG} closer to
  the brain: Performance of on-scalp sensor arrays. Neuroimage 147, 542--553.

\bibitem[{Iivanainen et~al.(2019)Iivanainen, Zetter, Gr\"an, Hakkarainen, and
  Parkkonen}]{IIVANAINEN2019}
Iivanainen, J., Zetter, R., Gr\"an, M., Hakkarainen, K., Parkkonen, L., 2019.
  On-scalp {MEG} system utilizing an actively shielded array of
  optically-pumped magnetometers. NeuroImage 194, 244--258.

\bibitem[{Iivanainen et~al.(2020)Iivanainen, Zetter, and
  Parkkonen}]{Iivanainen2020}
Iivanainen, J., Zetter, R., Parkkonen, L., 2020. Potential of on-scalp meg:
  Robust detection of human visual gamma-band responses. Human Brain Mapping
  41~(1), 150--161.

\bibitem[{Jackson(1998)}]{Jackson}
Jackson, J.~D., 1998. {Classical electrodynamics; 3rd ed.} Wiley, New York, NY.

\bibitem[{Jobst et~al.(2020)Jobst, Bartolomei, Diehl, Frauscher, Kahane,
  Minotti, Sharan, Tardy, Worrell, and Gotman}]{Jost2020}
Jobst, B.~C., Bartolomei, F., Diehl, B., Frauscher, B., Kahane, P., Minotti,
  L., Sharan, A., Tardy, N., Worrell, G., Gotman, J., 2020. {Intracranial EEG
  in the 21st century}. Epilepsy Currents 20~(4), 180--188.

\bibitem[{Labyt et~al.(2019)Labyt, Corsi, Fourcault, Palacios~Laloy, Bertrand,
  Lenouvel, Cauffet, Le~Prado, Berger, and Morales}]{Labyt2019}
Labyt, E., Corsi, M., Fourcault, W., Palacios~Laloy, A., Bertrand, F.,
  Lenouvel, F., Cauffet, G., Le~Prado, M., Berger, F., Morales, S., 2019.
  Magnetoencephalography with optically pumped {4He} magnetometers at ambient
  temperature. IEEE Trans. Med. Imaging 38, 90--98.

\bibitem[{Larson and Taulu(2018)}]{Larson2018}
Larson, E., Taulu, S., 2018. {Reducing sensor noise in MEG and EEG recordings
  using oversampled temporal projection}. IEEE Transactions on Biomedical
  Engineering 65~(5), 1002--1013.

\bibitem[{{Lucena G\'omez} et~al.(2021){Lucena G\'omez}, Peigneux, Wens, and
  Bourguignon}]{LucenaGomez2021}
{Lucena G\'omez}, G., Peigneux, P., Wens, V., Bourguignon, M., 2021.
  Localization accuracy of a common beamformer for the comparison of two
  conditions. NeuroImage 230, 117793.

\bibitem[{Marhl et~al.(2022)Marhl, Sander, and Jazbinsek}]{Marhl2022}
Marhl, U., Sander, T., Jazbinsek, V., 2022. Simulation study of different
  opm-meg measurement components. Sensors 22~(9).

\bibitem[{Mellor et~al.(2022)Mellor, Tierney, O'Neill, Alexander, Seymour,
  Holmes, López, Hill, Boto, Rea, Roberts, Leggett, Bowtell, Brookes, Maguire,
  Walker, and Barnes}]{Mellor2022}
Mellor, S., Tierney, T.~M., O'Neill, G.~C., Alexander, N., Seymour, R.~A.,
  Holmes, N., López, J.~D., Hill, R.~M., Boto, E., Rea, M., Roberts, G.,
  Leggett, J., Bowtell, R., Brookes, M.~J., Maguire, E.~A., Walker, M.~C.,
  Barnes, G.~R., 2022. {Magnetic field mapping and correction for moving
  OP-MEG}. IEEE Transactions on Biomedical Engineering 69~(2), 528--536.

\bibitem[{Muravchik and Nehorai(2001)}]{Muravchik2001}
Muravchik, C.~H., Nehorai, A., 2001. {EEG/MEG error bounds for a static dipole
  source with a realistic head model.} IEEE Transactions on Signal Processing
  49~(3), 470--484.

\bibitem[{Nardelli et~al.(2020)Nardelli, Perry, Krzyzewski, and
  Knappe}]{Nardelli2020}
Nardelli, N., Perry, A., Krzyzewski, S., Knappe, S., 2020. A conformal array of
  microfabricated optically-pumped first-order gradiometers for
  magnetoencephalography. EPJ Quantum Technol. 7, 11.

\bibitem[{Nardelli et~al.(2019)Nardelli, Krzyzewski, and
  Knappe}]{Nardelli2019b}
Nardelli, N.~V., Krzyzewski, S.~P., Knappe, S.~A., 2019. Reducing crosstalk in
  optically-pumped magnetometer arrays. Physics in Medicine and Biology
  64~(21), 21NT03.

\bibitem[{Nugent et~al.(2022)Nugent, {Benitez Andonegui}, Holroyd, and
  Robinson}]{Nugent2022}
Nugent, A.~C., {Benitez Andonegui}, A., Holroyd, T., Robinson, S.~E., 2022.
  On-scalp magnetocorticography with optically pumped magnetometers: Simulated
  performance in resolving simultaneous sources. Neuroimage: Reports 2~(2),
  100093.

\bibitem[{Sekihara et~al.(2005)Sekihara, Sahani, and Nagarajan}]{Sekihara2005}
Sekihara, K., Sahani, M., Nagarajan, S.~S., 2005. Localization bias and spatial
  resolution of adaptive and non-adaptive spatial filters for meg source
  reconstruction. NeuroImage 25~(4), 1056--1067.

\bibitem[{Seymour et~al.(2022)Seymour, Alexander, Mellor, O'Neill, Tierney,
  Barnes, and Maguire}]{Seymour2022}
Seymour, R.~A., Alexander, N., Mellor, S., O'Neill, G.~C., Tierney, T.~M.,
  Barnes, G.~R., Maguire, E.~A., 2022. Interference suppression techniques for
  opm-based meg: Opportunities and challenges. NeuroImage 247, 118834.

\bibitem[{Tarantola(2006)}]{Tarantola2006Nature}
Tarantola, A., 2006. {Popper, Bayes and the inverse problem}. Nature Physics
  2~(8), 492--494.

\bibitem[{Taulu et~al.(2004)Taulu, Kajola, and Simola}]{Taulu2004}
Taulu, S., Kajola, M., Simola, J., 2004. Suppression of interference and
  artifacts by the signal space separation method. Brain Topogr. 16, 269--275.

\bibitem[{Taulu et~al.(2005)Taulu, Simola, and Kajola}]{Taulu2005}
Taulu, S., Simola, J., Kajola, M., 2005. Applications of the signal space
  separation method. IEEE Trans. Sig. Proc. 53, 3359--3372.

\bibitem[{Tierney et~al.(2019)Tierney, Holmes, Mellor, L\'opez, Roberts, Hill,
  Boto, Leggett, Shah, Brookes, Bowtell, and Barnes}]{Tierney2019}
Tierney, T.~M., Holmes, N., Mellor, S., L\'opez, J.~D., Roberts, G., Hill,
  R.~M., Boto, E., Leggett, J., Shah, V., Brookes, M.~J., Bowtell, R., Barnes,
  G.~R., 2019. Optically pumped magnetometers: From quantum origins to
  multi-channel magnetoencephalography. NeuroImage 199, 598--608.

\bibitem[{Tierney et~al.(2021)Tierney, Levy, Barry, Meyer, Shigihara, Everatt,
  Mellor, Lopez, Bestmann, Holmes, Roberts, Hill, Boto, Leggett, Shah, Brookes,
  Bowtell, Maguire, and Barnes}]{Tierney2021}
Tierney, T.~M., Levy, A., Barry, D.~N., Meyer, S.~S., Shigihara, Y., Everatt,
  M., Mellor, S., Lopez, J.~D., Bestmann, S., Holmes, N., Roberts, G., Hill,
  R.~M., Boto, E., Leggett, J., Shah, V., Brookes, M.~J., Bowtell, R., Maguire,
  E.~A., Barnes, G.~R., 2021. Mouth magnetoencephalography: A unique
  perspective on the human hippocampus. NeuroImage 225, 117443.

\bibitem[{Tierney et~al.(2020)Tierney, Mellor, O'Neill, Holmes, Boto, Roberts,
  Hill, Leggett, Bowtell, Brookes, and Barnes}]{Tierney2020}
Tierney, T.~M., Mellor, S., O'Neill, G.~C., Holmes, N., Boto, E., Roberts, G.,
  Hill, R.~M., Leggett, J., Bowtell, R., Brookes, M.~J., Barnes, G.~R., 2020.
  {Pragmatic spatial sampling for wearable MEG arrays}. Scientific Reports 10,
  21609.

\bibitem[{Tierney et~al.(2022)Tierney, Mellor, O'Neill, Timms, and
  Barnes}]{Tierney2022}
Tierney, T.~M., Mellor, S., O'Neill, G.~C., Timms, R.~C., Barnes, G.~R., 2022.
  Spherical harmonic based noise rejection and neuronal sampling with
  multi-axis opms. NeuroImage 258, 119338.

\bibitem[{van Veen et~al.(1997)van Veen, van Drongelen, Yuchtman, and
  Suzuki}]{VanVeen1997}
van Veen, B.~D., van Drongelen, W., Yuchtman, M., Suzuki, A., 1997.
  Localization of brain electrical activity via linearly constrained minimum
  variance spatial filtering. IEEE Transactions on Biomedical Engineering 44,
  867--880.

\bibitem[{Vivekananda et~al.(2020)Vivekananda, Mellor, Tierney, Holmes, Boto,
  Leggett, Roberts, Hill, Litvak, Brookes, Bowtell, Barnes, and
  Walker}]{Vivekananda2020}
Vivekananda, U., Mellor, S., Tierney, T.~M., Holmes, N., Boto, E., Leggett, J.,
  Roberts, G., Hill, R.~M., Litvak, V., Brookes, M.~J., Bowtell, R., Barnes,
  G.~R., Walker, M.~C., 2020. Optically pumped magnetoencephalography in
  epilepsy. Annals of Clinical and Translational Neurology 7~(3), 397--401.

\bibitem[{Vrba et~al.(2004)Vrba, Robinson, and McCubbin}]{Vrba2004}
Vrba, J., Robinson, S.~E., McCubbin, J., 2004. How many channels are needed for
  {MEG}? Neurol Clin Neurophysiol 99.

\bibitem[{Wens et~al.(2015)Wens, Marty, Mary, Bourguignon, Op~de Beeck,
  Goldman, Van~Bogaert, Peigneux, and De~Ti{\`e}ge}]{Wens2015}
Wens, V., Marty, B., Mary, A., Bourguignon, M., Op~de Beeck, M., Goldman, S.,
  Van~Bogaert, P., Peigneux, P., De~Ti{\`e}ge, X., 2015. {A geometric
  correction scheme for spatial leakage effects in MEG/EEG seed-based
  functional connectivity mapping}. Hum Brain Mapp 36~(11), 4604--4621.

\bibitem[{Widjaja(2022)}]{Widjaja2022}
Widjaja, E., 2022. Wearable magnetoencephalography: Reality or science fiction?
  Radiology 304~(2), 435--436.

\bibitem[{Worsley et~al.(1996)Worsley, Marrett, Neelin, Vandal, Friston, and
  Evans}]{Worsley1996}
Worsley, K.~J., Marrett, S., Neelin, P., Vandal, A.~C., Friston, K.~J., Evans,
  A.~C., 1996. A unified statistical approach for determining significant
  signals in images of cerebral activation. Human Brain Mapping 4~(1), 58--73.

\bibitem[{Zangwill(2012)}]{Zangwill2012}
Zangwill, A., 2012. Modern Electrodynamics. Cambridge University Press.

\end{thebibliography}

\end{document}